\documentclass{mn2e}
\usepackage{graphicx}
\usepackage{subfigure}
\usepackage{amsmath}

\newcommand{\etal}{et~al.}

\newcommand{\Msun}{\mbox{${\rm M}_{\odot}$}}
\newcommand{\Mstar}{\mbox{${\rm M}_{\star}$}}

\begin{document}

\title[Magnetic fields in galaxies]{Magnetic fields in galaxies: I. Radio disks in local late-type galaxies.}

\author[Shabala, Mead \& Alexander]{S.S. Shabala$^{\star,1}$, J.M.G. Mead$^2$ and P. Alexander$^{3,4}$\\
$^1$ Oxford Astrophysics, Denys Wilkinson Building, Keble Road, Oxford OX1 3RH, United Kingdom\\
$^2$ Institute of Astronomy, Madingley Road, Cambridge CB3 0HA, United Kingdom\\
$^3$ Astrophysics Group, Cavendish Laboratory, J.J. Thomson Avenue, Cambridge CB3 0HE, United Kingdom\\
$^4$ Kavli Institute for Cosmology, Madingley Road, Cambridge CB3 0HA, United Kingdom\\
Email: stas.shabala@astro.ox.ac.uk}

\maketitle

\begin{abstract}

We develop an analytical model to follow the cosmological evolution of magnetic fields in disk galaxies. Our assumption is that fields are amplified from a small seed field via magnetohydrodynamical (MHD) turbulence. We further assume that this process is fast compared to other relevant timescales, and occurs principally in the cold disk gas. We follow the turbulent energy density using the Shabala \& Alexander (2009) galaxy formation and evolution model. Three processes are important to the turbulent energy budget: infall of cool gas onto the disk and supernova feedback increase the turbulence; while star formation removes gas and hence turbulent energy from the cold gas. Finally, we assume that field energy is continuously transferred from the incoherent random field into an ordered field by differential galactic rotation.

Model predictions are compared with observations of local late type galaxies by Fitt \& Alexander (1993) and Shabala \etal\/ (2008). The model reproduces observed magnetic field strengths and luminosities in low and intermediate-mass galaxies. These quantities are overpredicted in the most massive hosts, suggesting that inclusion of gas ejection by powerful AGNs is necessary in order to quench gas cooling and reconcile the predicted and observed magnetic field strengths.

\end{abstract}


\section{Introduction}
\label{sec:introduction}

Radio synchrotron emission of high energy electrons in the interstellar medium (ISM) indicates the presence of magnetic fields in galaxies. Rotation measures (RM) of background polarized sources indicate two varieties of field: a random field, which is not coherent on scales larger than the turbulence of the ISM; and a spiral ordered field which exhibits large-scale coherence (e.g. Stepanov \etal\/ 2008). For a typical galaxy these fields have strengths of a few $\mu$G. In a galaxy such as M\,51, the coherent magnetic field is observed to be associated with the optical spiral arms \cite{PatrikeevEA06}. Such fields are important in star formation and the physics of cosmic rays, and could also have an effect on galaxy evolution, yet, despite their importance, questions about their origin, evolution and structure remain largely unsolved. 

The Square Kilometre Array (SKA) will help us answer questions such as these. The SKA will observe polarized synchrotron emission \cite{GaenslerEA04}, and the ``All-Sky SKA Rotation Measure Survey'' will expand RM data sets by five orders of magnitude, mapping the magnetic fields of galaxies in unprecedented detail \cite{StepanovEA08,Gaensler06}. In particular, the SKA will provide data on the evolution of galactic magnetic fields to high redshift \cite{Gaensler06}, making a theoretical model for this process very valuable. Here we present such a model within the context of hierarchical structure formation.

Our model is based on the observed properties of galactic magnetic fields. Observations and simulations suggest that the random field is generated by turbulence in the ISM, which is modeled as a single-phase magnetohydrodynamic (MHD) fluid, within which magnetic field lines are frozen (e.g. Cho \etal\/ 2009). Simulations have shown that in such a turbulent MHD fluid, the random magnetic field energy and turbulent fluid kinetic energy are approximately equal after a few turnover times \cite{ChoEA09}, so by writing an equation for the turbulent energy in the ISM the random magnetic field energy can also be evaluated. The large-scale ordered field is produced by the differential rotation of the galaxy, which winds the random field into a spiral - this is the basic operation of a dynamo. A simple model for this process is postulated. The energy sources which contribute or remove turbulence from the ISM are supernovae, star formation and accretion of gas from the hot gas halo, and these parameters can be determined from well-established semi-analytic galaxy models, in which galaxies form from gas condensing at the centres of hierarchically merging haloes \cite{WhiteRees78,LaceySilk91,CrotonEA06}. The formation of magnetic fields in galaxies is intimately linked to galaxy formation, since the same physical processes are at work in both cases.

Recently, Arshakian \etal\/ (2009) have presented a qualitatively similar model for the evolution of magnetic fields in late-type galaxies. These authors considered three processes. At high redshift, the initial seed field of $\sim 10^{-18}$~Gauss is amplified via the Biermann battery mechanism to generate what they refer to as the regular field. This field is rapidly (on eddy turnover timescale, $\sim 10^7$~years for the Milky Way) amplified by virialization turbulence. The mean-field galactic dynamo mechanism then amplifies this regular seed field, with a typical $e$-folding time of $\sim 10^8$ years, until it reaches equilibrium with turbulent dissipation. By contrast, given the short (compared to cosmological) timescales involved, in our model we simply assume instantaneous equality between the magnetic field energy and turbulent kinetic energy to estimate what we refer to as the random field. This field is then ordered by the mean-scale dynamo on much longer timescales.

While our treatment of large-scale dynamo generation is simplified compared to that of Arshakian \etal\/, it encapsulates the relevant physics and has the advantage of being simple enough to be included in a cosmological framework. In that sense this work is highly complementary to the Arshakian \etal\/ (2009) study. We present here the first (to our knowledge) attempt to include magnetic fields in a self-consistent galaxy formation and evolution model. A number of galaxy properties are predicted, and we compare these with available data. This paper primarily focuses on the radio properties of local late-type galaxies.

The paper is organised as follows. In Section~\ref{sec:model} we outline the adopted galaxy formation model, and develop models for both the random and ordered magnetic fields. Section~\ref{sec:parameterSpace} investigates the salient features of our model and constrains various parameters. Predictions of our model are compared with available observational data in Sections~\ref{sec:BfieldVsStellarMass} and \ref{sec:localRLF}. We summarise our findings in Section~\ref{sec:conclusions}.

Throughout the paper, we adopt a flat cosmology of $\Omega_{\rm M}=0.3$, $\Omega_\Lambda=0.7$, $h=0.7$ and $\sigma_8 = 0.9$, consistent with the 2dFGRS \cite{CollessEA01} and WMAP \cite{SeljakEA05} results.

\section{Tracing galactic magnetic fields}
\label{sec:model}

\subsection{Galaxy evolution model}
\label{sec:galEvModel}

We employ the analytical galaxy formation and evolution model of Shabala \& Alexander (2009). This traces gas cooling, star formation, and various feedback processes in a cosmological context. The model simultaneously reproduces the local galaxy properties, star formation history of the Universe, the evolution of the stellar mass function to $z \sim 1.5$, and the early build-up of massive galaxies. While details can be found in Shabala \& Alexander (2009), below we briefly outline those features of the model pertinent to the current paper.

The model follows {\it average} properties of a galaxy of a given mass at redshift zero. Mass accretion history of the host halo is followed via the formalism introduced by van den Bosch (2002), providing analytical fits to detailed N-body simulations. Following the conventional picture, baryons are assumed to be accreted onto the halo together with the dark matter. Thus, in our model each halo with a given redshift-zero mass can only host one galaxy. Reionization feedback from first generations of structure limits the accreted baryonic fraction. The accreted gas is shocked to the virial temperature and density, and cools radiatively. Once cold, the gas is added to the galaxy disk on a dynamical timescale, where star formation can take place.

These cooling processes complete with various modes of feedback. Stars more massive than $8$~\Msun\/ inject supernova shocks into the surrounding gas as they expire; this type of feedback is important in low and intermediate mass galaxies. Gas cooling also fuels intermittent AGN activity, which becomes important in haloes more massive than $10^{12}$~\Msun\/.

Such detailed treatment of the gas heating and cooling processes as well as star formation in an analytical framework makes this model ideal for following the cosmological evolution of galactic magnetic fields, as we outline below.

\subsection{Magnetic fields in galaxy disks}
\label{sec:turbAsProxy}
Observations of spiral galaxies, for instance using faraday rotation measures, have indicated the presence of random and ordered magnetic fields, of magnitudes of a few $\mu$G (e.g. Stepanov \etal\/ 2008). The random fields are coherent on small scales, of order parsecs. The ordered fields are coherent on much larger scales, and are observed to follow the spiral arms of the galaxy. In this section we describe how we model both types of field, and show how the random and ordered fields are linked by the differential rotation of the galaxy.

Simulations of a turbulent MHD fluid suggest that random magnetic fields can be generated within such fluids via turbulence \cite{ChoEA09,ChoVishniac00,KulsrudEA97}. These simulations have shown that in such a turbulent MHD fluid, equality between random magnetic field energy and turbulent fluid kinetic energy is reached after several turnover times. This result forms the basis for our model of random magnetic fields in galaxies. We then model the formation of the large-scale ordered field by a simplified model based on the differential rotation of the galaxy.

The mechanism by which turbulence generates a microgauss random magnetic field is field line stretching. The turbulent fluid motions draw out field lines, and thus, in ideal MHD, the field strength along that direction must increase. This increases magnetic pressure, which opposes the stretching process. On a timescale of several eddy turnover times, this build up in magnetic pressure suppresses further field line stretching, and the process ceases when the random magnetic field energy is equal to the energy in turbulence of the fluid (Cho \& Vishniac 2000). 

Whilst the random field can be accounted for through turbulence in the ISM, a mechanism for generating the ordered field must invoke the differential rotation of the galaxy. The galactic dynamo theory is commonly used to explain the presence of large-scale coherent ordered magnetic fields in spiral galaxies (e.g. Ruzmaikin \etal\/ 1988). In dynamo theory, a radial magnetic field is wound into a toroidal field, and amplified, via the differential rotation of the galaxy. In addition to this so-called ``$\omega$-effect'', a further process exists whereby a toroidal field is transformed into a radial field - this is known as the ``$\alpha$-effect'' \cite{Parker55}. Arshakian \etal\/ (2009) have presented detailed calculations of these processes. In the present paper, a simpler model for the generation of an ordered field is used. The generation of magnetic fields is broken up into a two step process. In the first step, the turbulence in the ISM stretches and amplifies the random magnetic field. In the second step, differential rotation winds the random field to create an ordered field. Such a division is physically motivated by the relevant timescales for the two steps. The random field is amplified on a timescale comparable to the eddie turnover time. This scale is much shorter than the characteristic timescale for ordered field generation, the galactic rotational period. In other words, it is a good approximation to treat the random field energy as amplified to a point at which it is equal to half the energy in turbulence, before allowing this random field to be converted into an ordered field by the large scale differential rotation. Such an approach has the advantage of being simple while capturing the relevant physics, and therefore can be implemented in a self-consistent galaxy formation and evolution model.

The ordering of the magnetic field, performed by the mean-scale dynamo, is a much slower process. Arshakian \etal\/ find the relevant timescale to be proportional to disk size. By contrast, in our approximation the $e$-folding time for this process is simply related to the angular rotation velocity of the galaxy, and is thus constant at a given redshift. It is worth pointing out that the two timescales are similar: Arshakian \etal\/ (their Equation 7) give the ordering timescale $t^*=\frac{h}{\Omega l}=\frac{t_{\rm rot}}{2 \pi} \left( \frac{h}{l} \right)$ where the ratio between disk scale height and turbulence length scale is $\frac{h}{l} \approx 5$, and angular velocity $\Omega$ of the galaxy can be expressed in terms of the rotation period $t_{\rm rot}$. Therefore, their $t^*$ is very close to our assumed rotation timescale $t_{\rm rot}$.

\subsubsection{Random fields}
\label{sec:Brandom}

As discussed above, since the turnover timescale of turbulent eddies is much less than cosmological timescales, the equality between random magnetic field energy and turbulent kinetic energy is assumed to hold instantaneously, at the point of energy injection. In the absence of an ordered field, the instantaneous equality between random field energy and turbulent energy demands half the energy injected into the ISM goes into turbulent fluid motion, and the other half into random magnetic field energy. The processes that inject and remove energy from the ISM will now be considered.

One of the most important sources of energy injection into the ISM are supernovae. They inject energy at a rate $\Psi_{\rm SF}e_{\rm SN}$, where $\Psi_{\rm SF}$ is the star formation rate in $M_\odot \mathrm{yr^{-1}}$ and $e_{\rm SN}$ is the energy released by supernovae per solar mass of star formation. The quantity $e_{\rm SN}$ can be estimated as follows. Assuming the minimum stellar mass to be 0.1$M_\odot$, and taking all stars with masses greater than 8$M_\odot$ to terminate as supernovae, the supernova rate $R_{\rm SN}$ can be related to the star formation rate by

\begin{eqnarray}
  R_{\rm SN} & = & \Psi_{\rm SF}\left(t\right)\frac{\int_{8}^{\infty}\phi(M)dM}{\int_{0.1}^{\infty}\phi(M)MdM} \nonumber\\
  & = & \Psi_{\rm SF}\left(t\right)\frac{\int_{8}^{\infty}M^{-2.35}dM}{\int_{0.1}^{\infty}M^{-2.35}MdM} ,
\end{eqnarray}
where we have assumed a Salpeter initial mass function $\phi(M)\propto M^{-2.35}$.

If each supernova injects energy $E_{\rm K}$ into the ISM,

\begin{equation}
  e_{\rm SN}=\frac{R_{\rm SN}E_{\rm K}}{\Psi_{\rm SF}} .
\end{equation}

If $E_{\rm K}\sim10^{44}\mathrm{J}$, approximately the kinetic energy of the supernova envelope, then $e_{\rm SN} \sim 10^{41}$~J\,$\Msun^{-1}$.

Star formation removes turbulent energy. If $E_{\rm turb}$ is the turbulent ISM energy, and $M_{\rm cold}$ the mass of cold gas in the galaxy, the turbulent energy removed by the formation of a star of mass $M_{\star}$ is $\frac{M_{\star}}{M_{\rm cold}} E_{\rm turb}$ (i.e. the fraction of total gas mass removed by star formation, multiplied by the turbulent energy). So the rate at which energy is removed by star formation is $\frac{\Psi_{\rm SF}}{M_{\rm cold}}E_{\rm turb}$.

Gas accreting from the dark matter halo deposits its potential energy in turbulence. If a gas packet of mass $\Delta M_{\rm cool}$ falls onto the galactic disk from a radius $r$, it adds energy $\int^{r_i} \frac{G M \left(<r_{\rm i} \right) \Delta M_{\rm cool}}{r^2} dr$ to the cold gas in the disk; we will assume all this energy goes into turbulence. If we sum over all such gas packets, and then differentiate with respect to time, we find that the energy injection rate from accretion is $\Sigma_{\rm i} \left( \int^{r_i} \frac{G M \left( <r \right) \dot{M}_{\rm cool, i}}{r^2} dr  \right)$.   

The total rate at which energy is injected into turbulence is equal to the sum of the three rates above. In front of each term in the equation we add factors which reflect the efficiency with which energy is added or removed from the ISM by a particular process. For the accretion this factor is labelled $\epsilon_{\rm grav}$, for supernovae feedback $\epsilon_{\rm SNe}$ and for star formation $\epsilon_{\rm SF}$. Half of the energy injected into the ISM goes into turbulent motions of the fluid, and the other half into the random magnetic field energy $E_{\rm B}^{\rm{ran}}$, so the equation for the evolution of the random field energy in the absence of differential galactic rotation is

\begin{eqnarray}
        \frac{dE_{\rm{B}}^{\rm{ran}}}{dt} = \frac{1}{2} \left[ \epsilon_{\rm SNe} \Psi_{\rm SF} {e_{\rm SN}} - \epsilon_{\rm SF} \frac{E_{\rm turb}}{M_{\rm cold}} \Psi_{\rm SF} \right. \nonumber \\
          + \left. \epsilon_{\rm grav} \Sigma_{\rm i} \left( \int^{r_i} \frac{G M(<r) \dot{M}_{\rm cool, i}}{r^2} dr  \right) \right]
\label{eqn:dUdt_random}
\end{eqnarray}
where the sum is over the gas parcels cooling at the timestep of interest. Here, $\Psi_{\rm SF}$, $M_{\rm cold}$ and $\dot{M}_{\rm cool}$ are calculated self-consistently from the galaxy formation model. Note that in the absence of rotation (i.e. when the whole field is in the random component) we have $\frac{dE_{\rm{B}}^{\rm{ran}}}{dt} = \frac{dE_{\rm{turb}}}{dt}$.

\subsubsection{Ordered fields}
\label{sec:Bordered}

As discussed above, we assume that the ordered field is created through galactic differential rotation. If at time $t=0$ the energy in the random field is $E_{\rm B, 0}^{\rm{ran}}$, assuming that there is no further energy injection, the energy in the ordered field at time $t$, $E_{\rm{B}}^{\rm{ord}}(t)$, is

\begin{equation}
  E_{\rm B}^{\rm{ord}}(t)=E_{\rm B, 0}^{\rm{ran}} \left(1-e^{-t/\tau_{\rm ord}}\right)\:,
\label{eqn:Eordered}
\end{equation}

where the timescale for energy transfer from the random to ordered field is $\tau_{\rm ord}$. We expect $\tau_{\rm ord}$ to be comparable with $t_{\rm rot}$, the rotational period of the galaxy, and therefore we set $\tau_{\rm ord}$=$t_{\rm rot}$.

In reality, energy is constantly injected into the ISM (by the processes considered above). Therefore we must extend the above calculation to determine how the energy in the ordered magnetic field varies when energy is continually injected to the random magnetic field. Let the random field energy be $E_{\rm B}^{\rm ran}\left(t\right)$ at time $t$. At $t+\Delta t$ the energy is $E_{\rm B}^{\rm ran} \left( t+\Delta t \right) \approx E_{\rm B}^{\rm{ran}}(t)+\left( \frac{dE_{\rm B}^{\rm{ran}}(t')}{dt'} \right)_{t} \Delta t$ . Thus $\left( \frac{dE_{\rm B}^{\rm{ran}}(t')}{dt'} \right)_{t} \Delta t$ has been added to the random field energy in the infinitesimal time interval $\Delta t$. For $\left( \frac{dE_{\rm B}^{\rm{ran}}(t')}{dt'} \right)_{t} <0$, the ordered field is assumed not to change - if the random energy declines there is no physical reason for the ordered field energy to decline also. If $\left( \frac{dE_{\rm B}^{\rm{ran}}(t')}{dt'} \right)_{t} >0$, using Equation~\ref{eqn:Eordered}, the ordered field energy at time $t_{0}$ due to the random field energy added in the time interval $\left[t, t+\Delta t \right]$ is,

\begin{equation}
  \Delta E_{\rm B}^{\rm{ord}} \left(t_{0}\right)=\left(\frac{dE_{\rm B}^{\rm{ran}}(t')}{dt'}\right)_{t}\Delta t\left(1-e^{\left(t-t_{0}\right)/t_{\rm rot}}\right)\:.
\end{equation}

The total energy in the ordered field can now be obtained by adding all contributions,

\begin{equation}
  E_{\rm B,ordered}\left(t_{0}\right)=\int_{0}^{t_{0}}\left(\frac{dE_{\rm B}^{\rm{ran}}(t')}{dt'}\right)_{t} f \left( t/t_{\rm rot} \right)dt\:,
\end{equation}

where,

\begin{equation}
  f \left( t / t_{\rm rot} \right)=\left\{\begin{aligned}
     1-e^{\left(t-t_{0}\right)/t_{\rm rot}} &\qquad \textrm{for } \frac{dE_{\rm B}^{\rm{ran}}}{dt}>0\\ 
    0 &\qquad \textrm{for } \frac{dE_{\rm B}^{\rm{ran}}}{dt} \leq 0
  \end{aligned}\;,\right.
\label{reducedshear}
\end{equation}

Equation \ref{eqn:dUdt_random} for the random field energy holds with an extra loss term included to reflect the fact that random field energy is continuously being converted into ordered field energy. The final equations for the evolution of the magnetic field energies are 

\begin{equation}
\frac{dE_{\rm B}^{\rm{ord}}}{dt}=\frac{dE_{\rm B}^{\rm{ran}}}{dt}f\left(t/t_{\rm rot}\right)
\label{eqn:du_ordered/dt}
\end{equation}

\begin{equation}
\frac{dE_{\rm B}^{\rm{ran}}}{dt}=\frac{1}{2}\frac{dE_{\rm{turb}}}{dt}-\frac{dE_{\rm B}^{\rm{ord}}}{dt} 
\label{eqn:du_rand/dt_final}
\end{equation}

\section{Preliminary results}
\label{sec:parameterSpace}

There are four free parameters in the model. Three parameters describe the efficiency with which turbulence is added to/removed from the cold gas by gravitational infall ($\epsilon_{\rm grav}$), supernovae feedback ($\epsilon_{\rm SNe}$) and star formation ($\epsilon_{\rm SF}$). These determine the total magnetic field strength. The fourth parameter is a timescale relating to the rate at which the random magnetic field is transferred to the ordered field, $\tau_{\rm{ord}}$, which in the previous section we have set to the rotation period of the galaxy.

\begin{figure*}
        \centering
	\subfigure[$M_{\rm halo} = 10^{11}$~\Msun\/, $\epsilon_{\rm grav}=0.003$]{\includegraphics[height=0.4\textwidth,clip,angle=270]{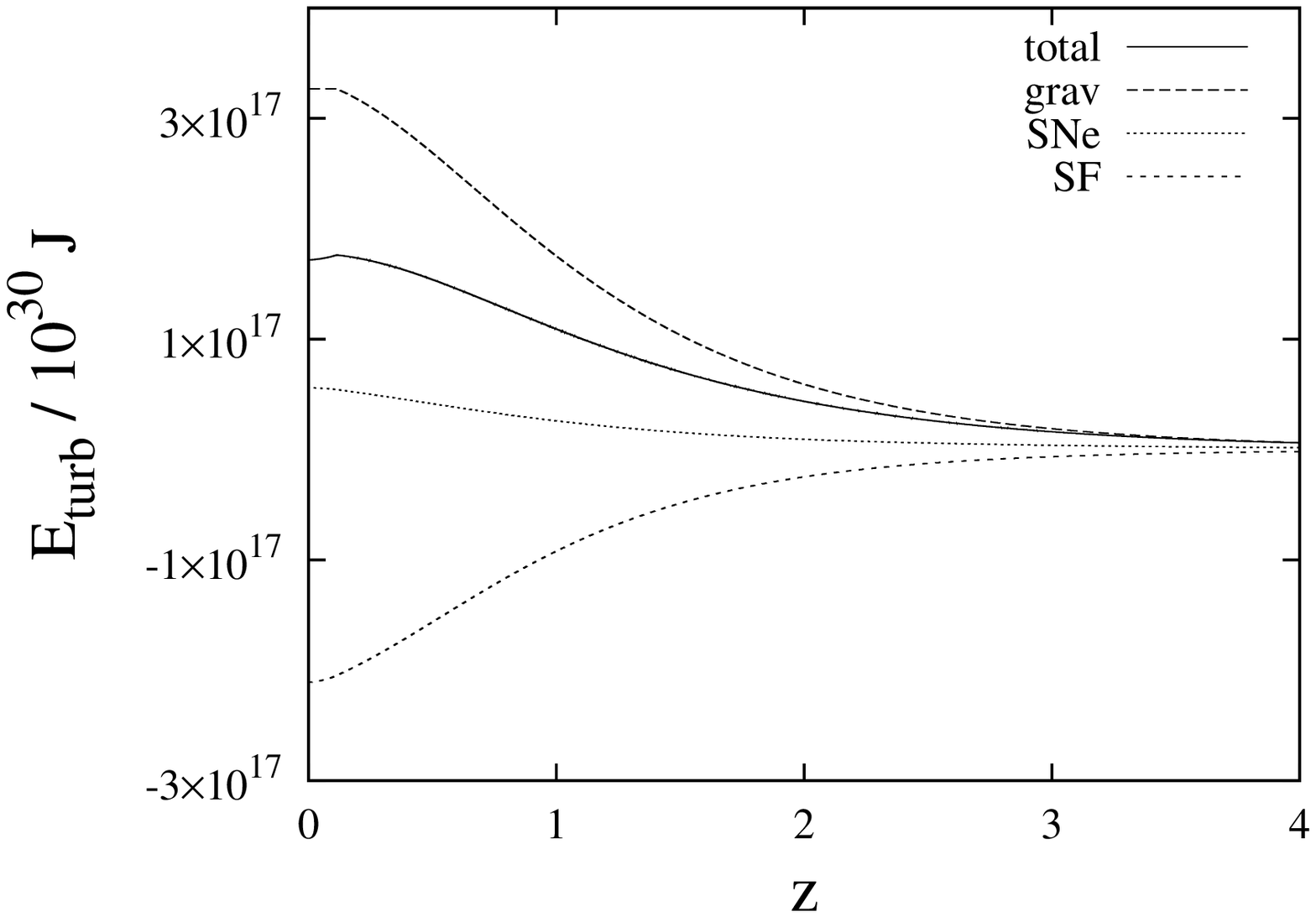}}
	\subfigure[$M_{\rm halo} = 10^{11}$~\Msun\/, $\epsilon_{\rm grav}=0.001$]{\includegraphics[height=0.4\textwidth,clip,angle=270]{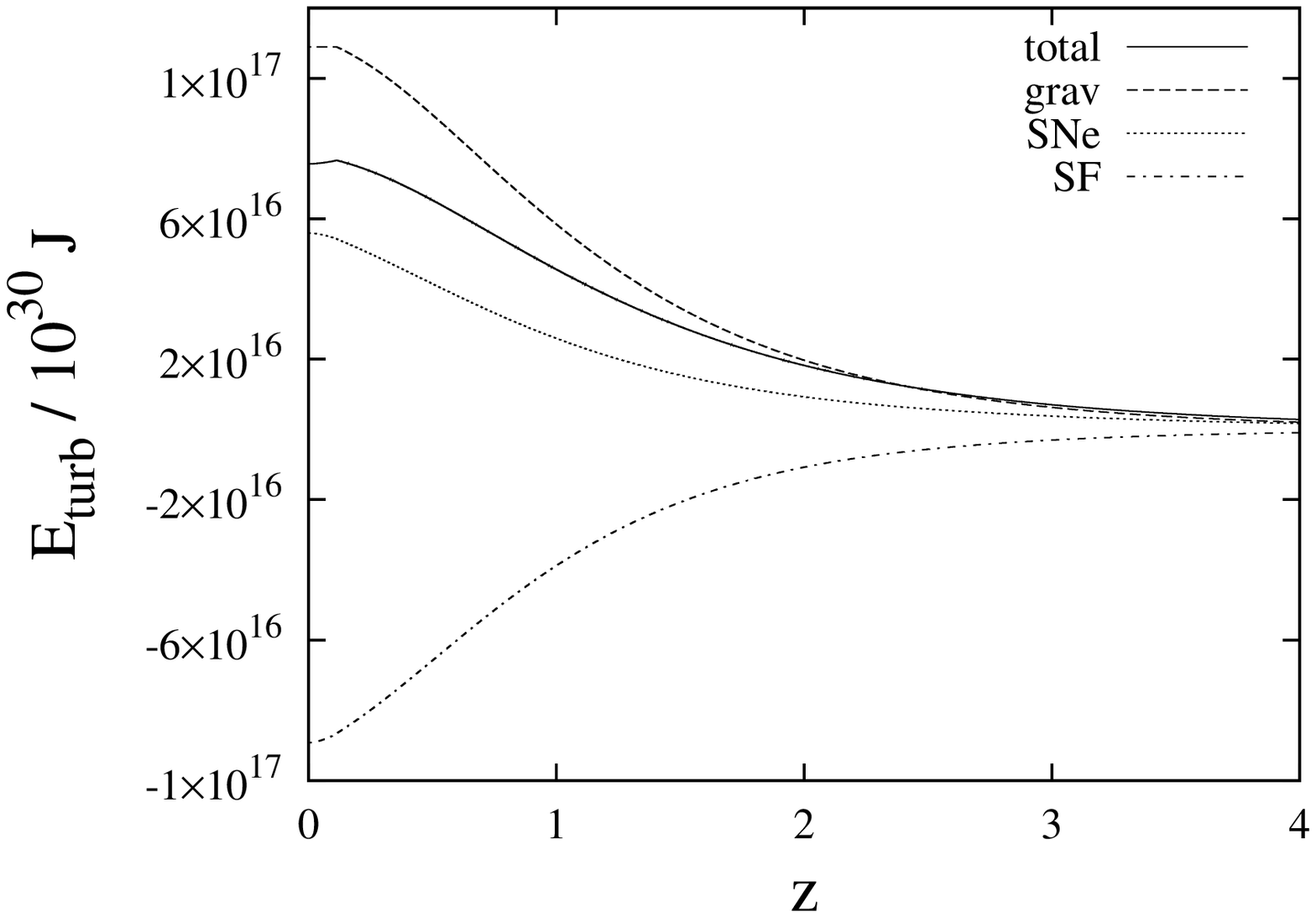}}
	\subfigure[$M_{\rm halo} = 10^{12}$~\Msun\/, $\epsilon_{\rm grav}=0.003$]{\includegraphics[height=0.4\textwidth,clip,angle=270]{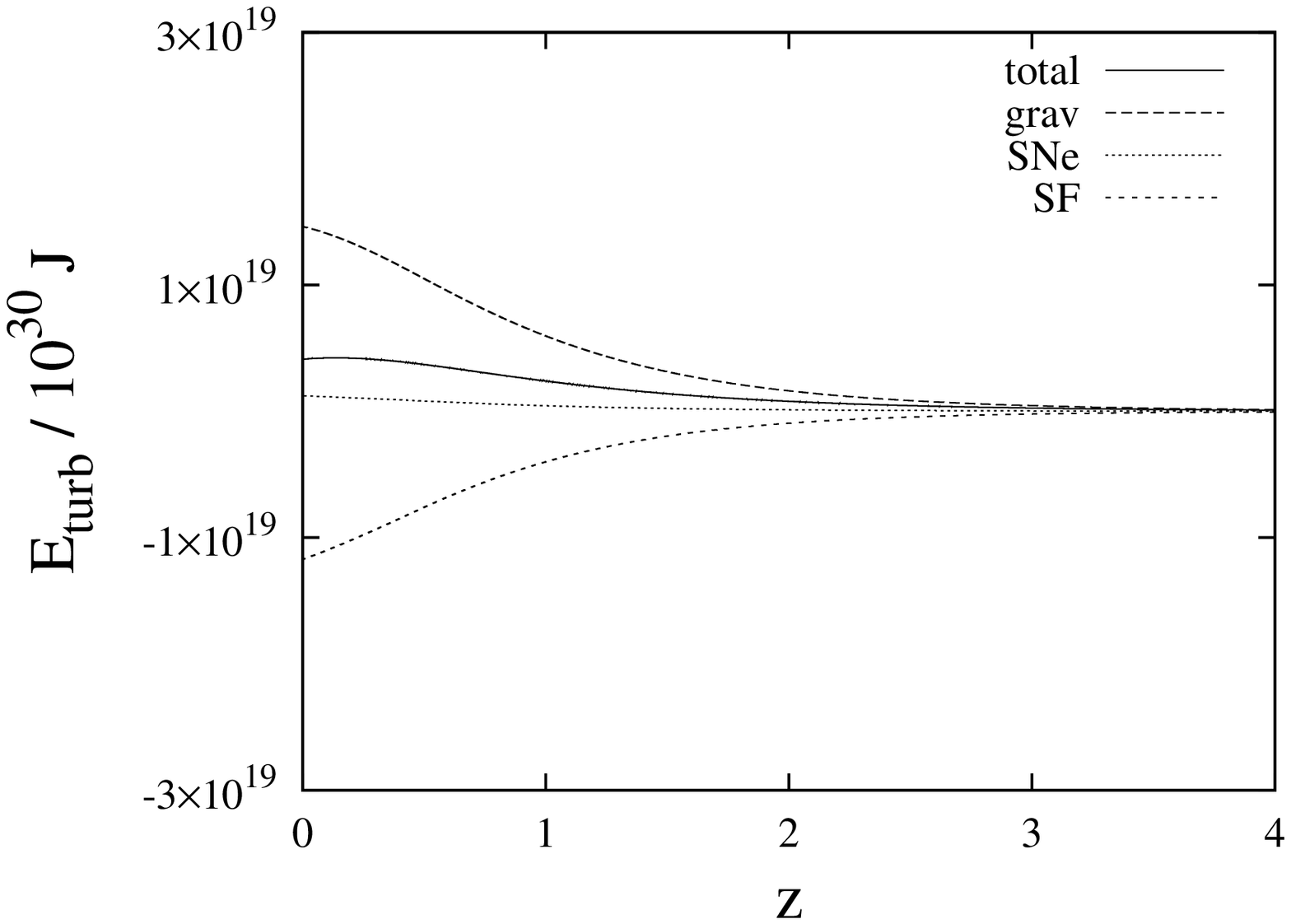}}
	\subfigure[$M_{\rm halo} = 10^{12}$~\Msun\/, $\epsilon_{\rm grav}=0.001$]{\includegraphics[height=0.4\textwidth,clip,angle=270]{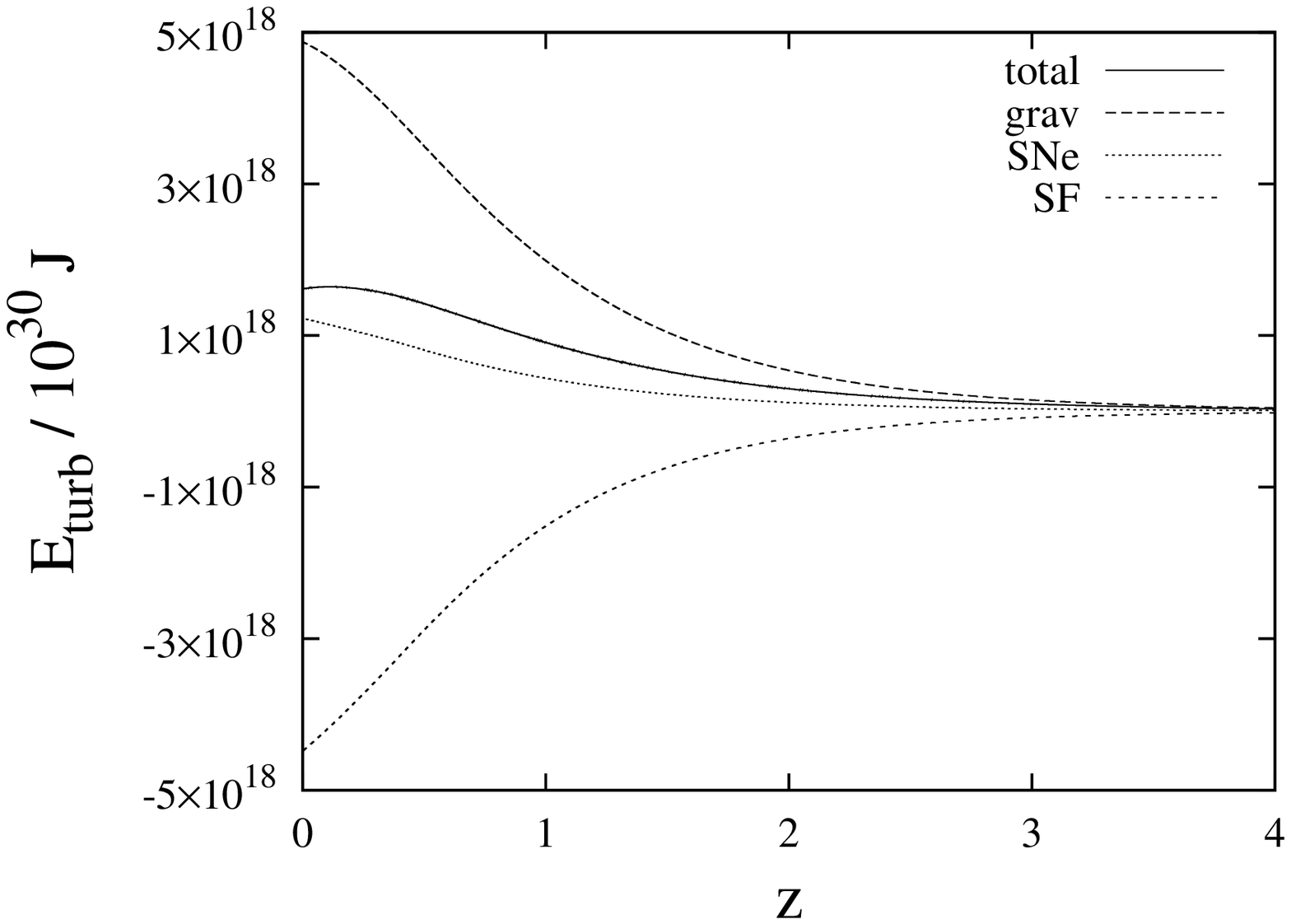}}
	\subfigure[$M_{\rm halo} = 10^{13}$~\Msun\/, $\epsilon_{\rm grav}=0.003$]{\includegraphics[height=0.4\textwidth,clip,angle=270]{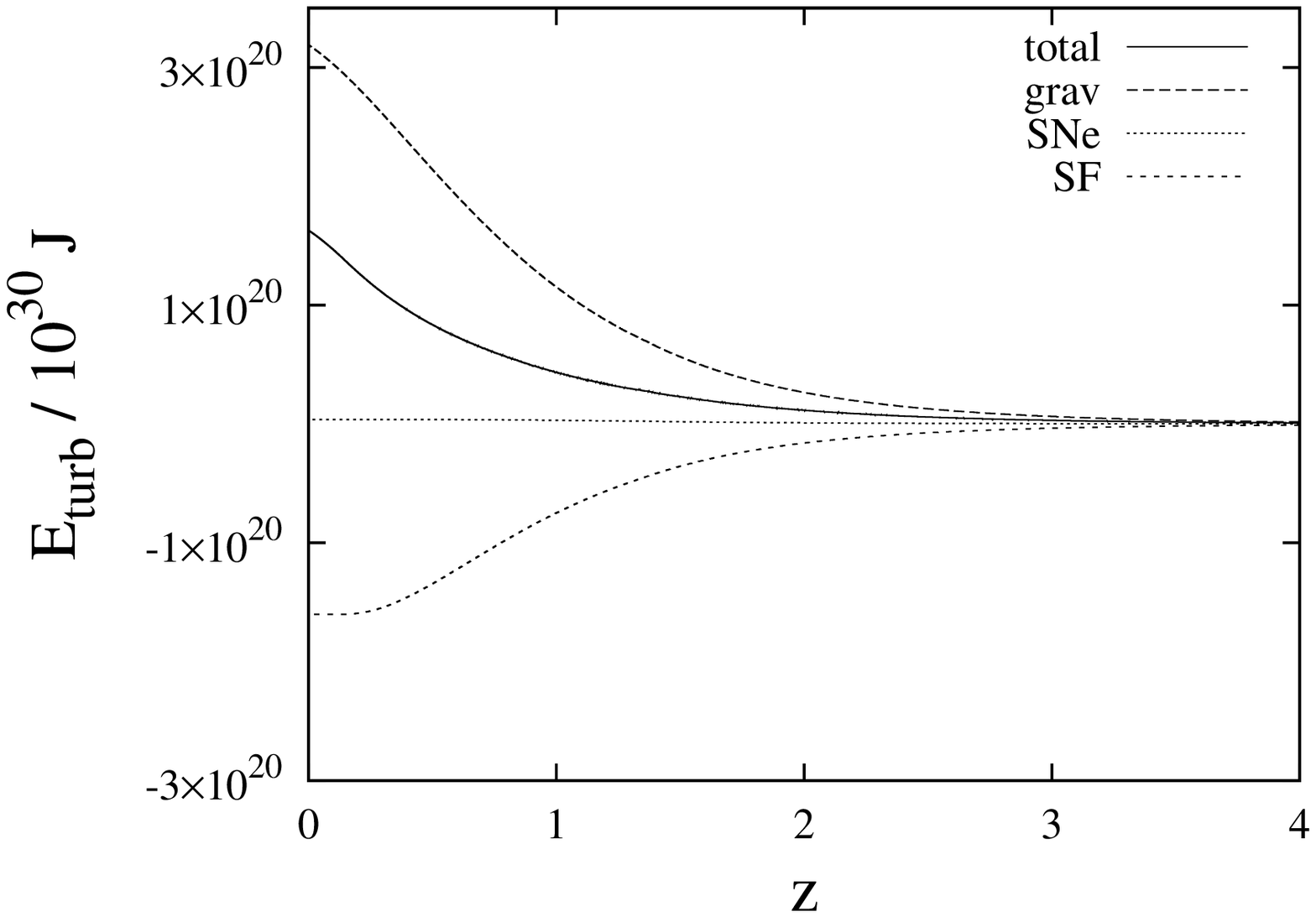}}
	\subfigure[$M_{\rm halo} = 10^{13}$~\Msun\/, $\epsilon_{\rm grav}=0.001$]{\includegraphics[height=0.4\textwidth,clip,angle=270]{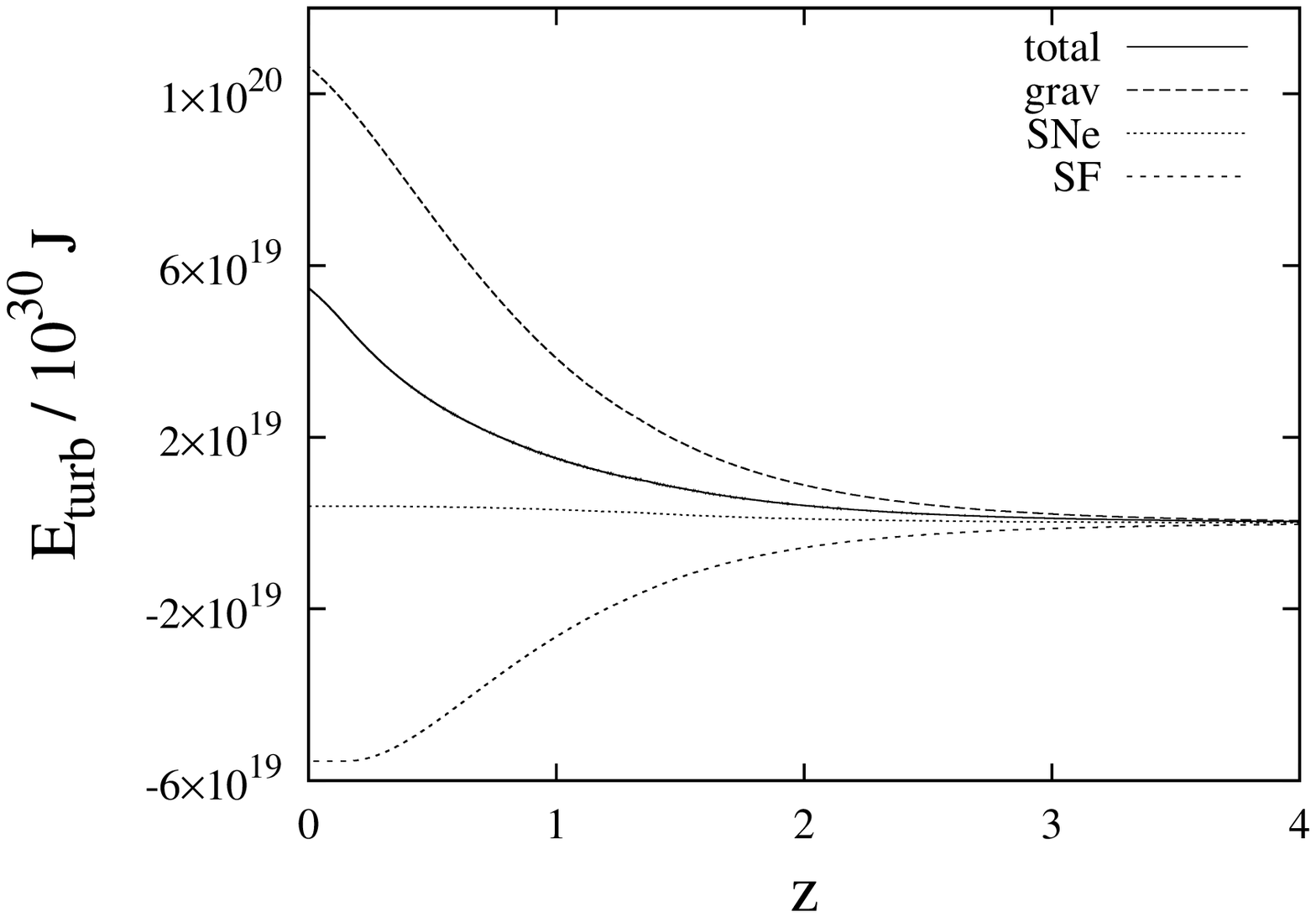}}
	\subfigure[$M_{\rm halo} = 10^{15}$~\Msun\/, $\epsilon_{\rm grav}=0.003$]{\includegraphics[height=0.4\textwidth,clip,angle=270]{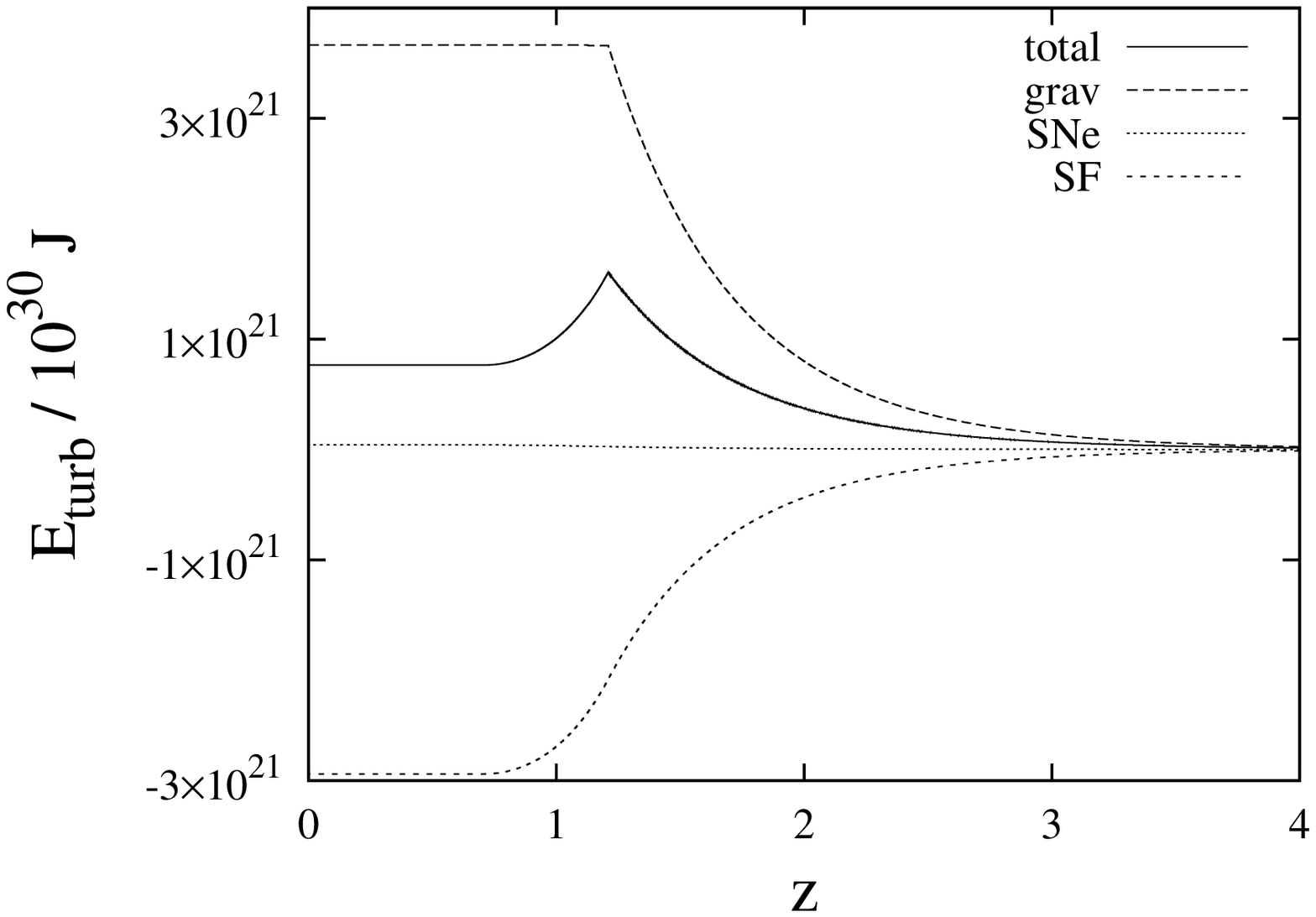}}
	\subfigure[$M_{\rm halo} = 10^{15}$~\Msun\/, $\epsilon_{\rm grav}=0.001$]{\includegraphics[height=0.4\textwidth,clip,angle=270]{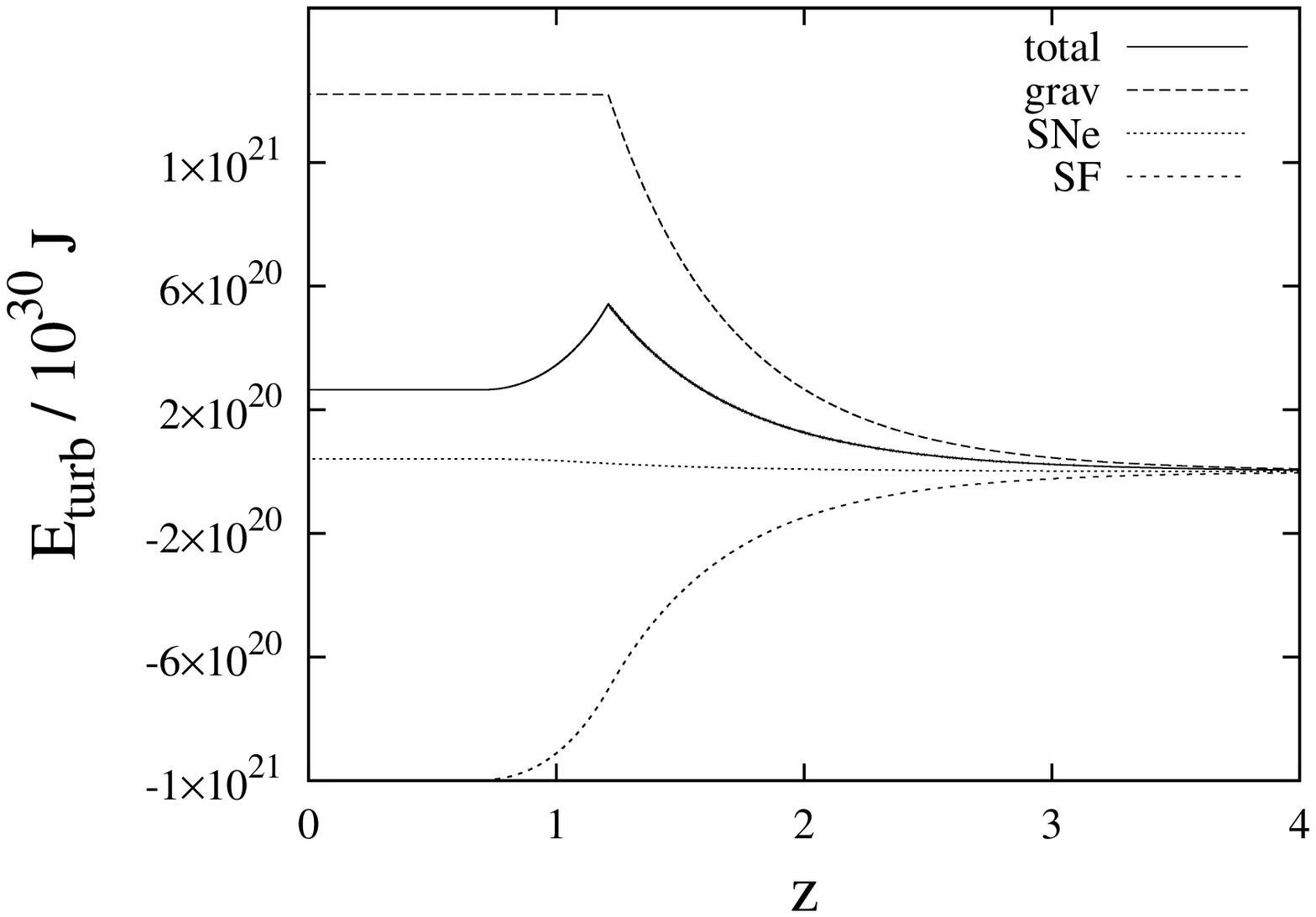}}
\caption{Total turbulent energy in the cold gas. {\it Left panels}: $\epsilon_{\rm grav}=0.003$. {\it Right panels}: $\epsilon_{\rm grav}=0.001$. All curves are for $\epsilon_{\rm SNe}=0.01$ and $\epsilon_{\rm SF}=1$. Star formation and cold gas infall dominate the SNe feedback contribution to turbulence. In high mass haloes, early truncation (at $z \sim 1$) of gas cooling, followed after a time lag by a truncation in star formation, results in the spike in, and the eventual flattening of, the turbulent energy curve.}
\label{fig:effPot} 
\end{figure*}

We set $\epsilon_{\rm SF}=1$, as physically one would expect all turbulent energy associated with the cold gas to be removed once the gas turns into stars. Figure~\ref{fig:effPot} shows the interplay between the three mechanisms with this value fixed for a range of halo masses. Shabala \& Alexander (2009) estimate that about 1 per cent of SNe energy must go towards heating and ejecting galactic gas in order to reproduce the low-mass end of the present day stellar mass function. We therefore expect $\epsilon_{\rm SNe}$ to be of this order, i.e. $\epsilon_{\rm SNe} \sim 0.01$. Figure~\ref{fig:effPot} shows that in this case gravitational infall and star formation dominate supernovae feedback, and thus the exact value of $\epsilon_{\rm SNe}$ is unimportant. We thus set $\epsilon_{\rm SNe} = 0.01$.

\begin{figure*}
        \centering
	\includegraphics[height=0.45\textwidth,clip,angle=270]{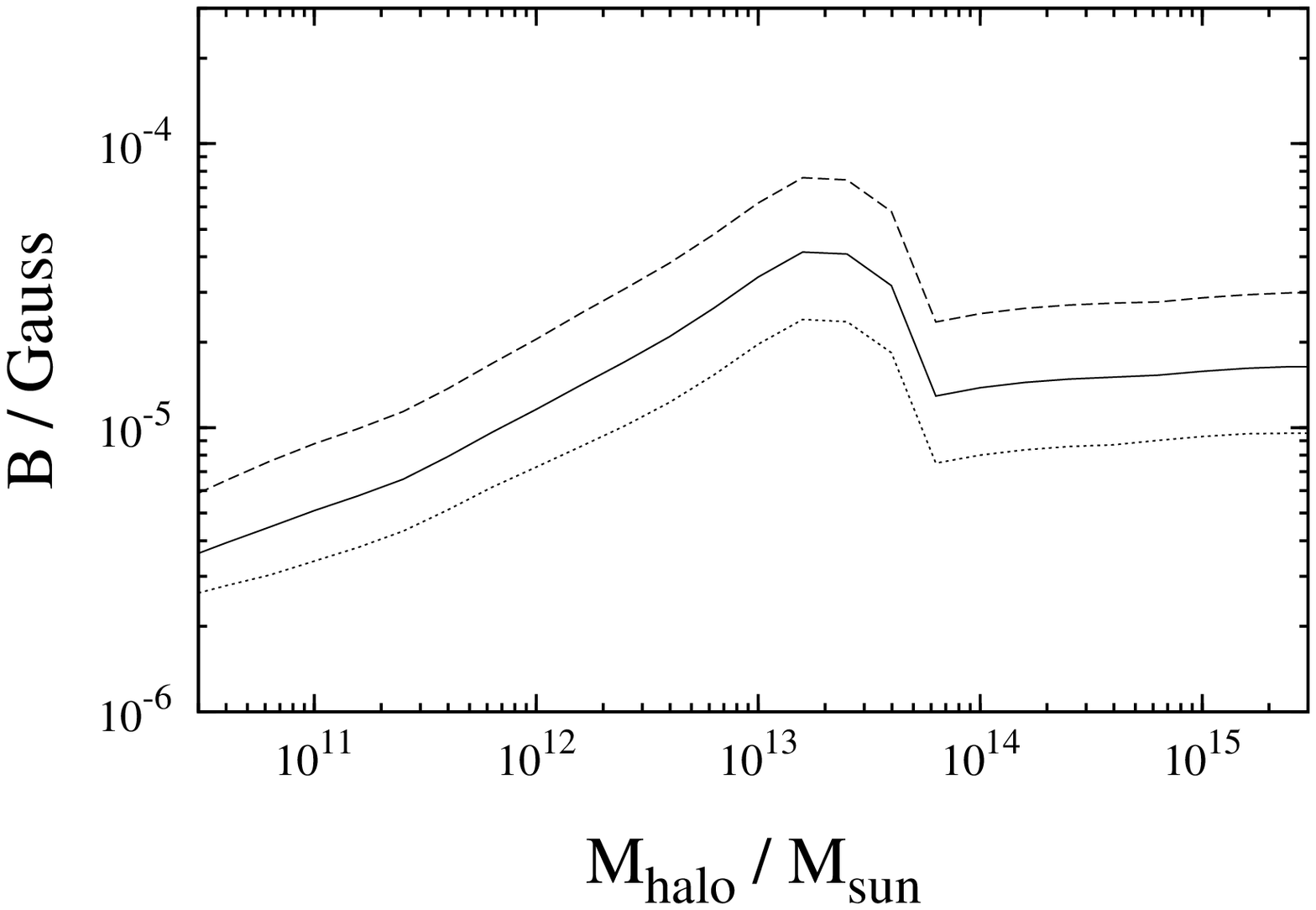}
\caption{Magnetic field strength at the present epoch as a function of halo mass for $\epsilon_{\rm grav}=0.003$ (solid line), $\epsilon_{\rm grav}=0.01$ (dashed) and $\epsilon_{\rm grav}=0.001$ (dotted line). All models have $\epsilon_{\rm SF}=1$, $\epsilon_{\rm SNe}=0.01$.}
\label{fig:BvsMhalo} 
\end{figure*}

As halo mass increases, so do the contributions to turbulence from gas cooling and the removal of turbulence due to star formation. In low mass haloes ($M_{\rm halo} \leq 10^{12}$~\Msun\/) at the present epoch, as halo mass increases, more turbulence is added through gas cooling than is removed by star formation. While significant amounts of gas can cool, an increase in the size of the galaxy disk means the gas surface density drops, and thus star formation is less efficient. As a result, the magnetic field strength increases with mass (Figure~\ref{fig:BvsMhalo}).

The turbulent energy density and magnetic field strength peak at $M_{\rm halo} \sim 10^{13}$~\Msun\/. At these masses, both supernovae and AGN feedback are too weak to quench large amounts of gas cooling. Massive haloes are more likely to host powerful AGNs \cite{SAAR08}, and thus in higher mass haloes (Figure~\ref{fig:effPot}g and h) AGN feedback truncates gas cooling and star formation at early epochs, with the turbulent energy density remaining unchanged since $z \sim 1$. Because the gas cooling time increases with decreasing density, powerful AGNs in massive hosts play a more important role in quenching gas cooling and star formation than do their lower power counterparts in less massive haloes. In the most massive haloes ($M_{\rm halo}>10^{15}$~\Msun\/) AGN feedback cannot completely quench gas cooling, and thus the magnetic field strength begins to increase once again.

The average magnetic field strength $B$ in a galactic disk is a function of the turbulent energy density. Thus, apart from the amount of turbulence present, another important parameter is the size of the disk. In the following section we use observations of local galaxies to constrain disk sizes, and consequently find the value of $\epsilon_{\rm grav} \sim 0.003$. This is the value adopted for the rest of the paper.

\begin{figure*}
        \centering
	\subfigure[$M_{\rm halo} = 10^{11}$~\Msun\/]{\includegraphics[height=0.45\textwidth,clip,angle=270]{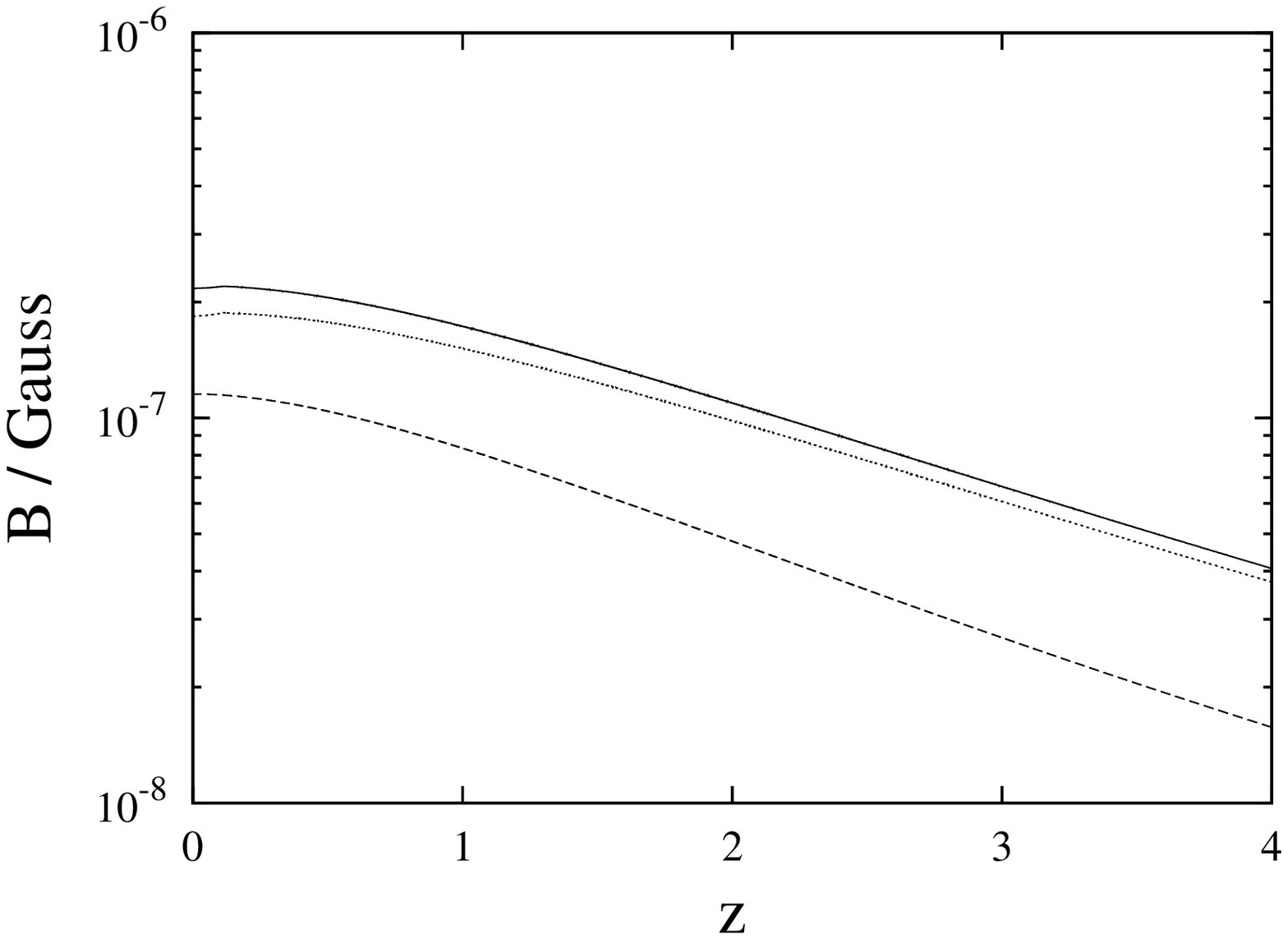}}
	\subfigure[$M_{\rm halo} = 10^{12}$~\Msun\/]{\includegraphics[height=0.45\textwidth,clip,angle=270]{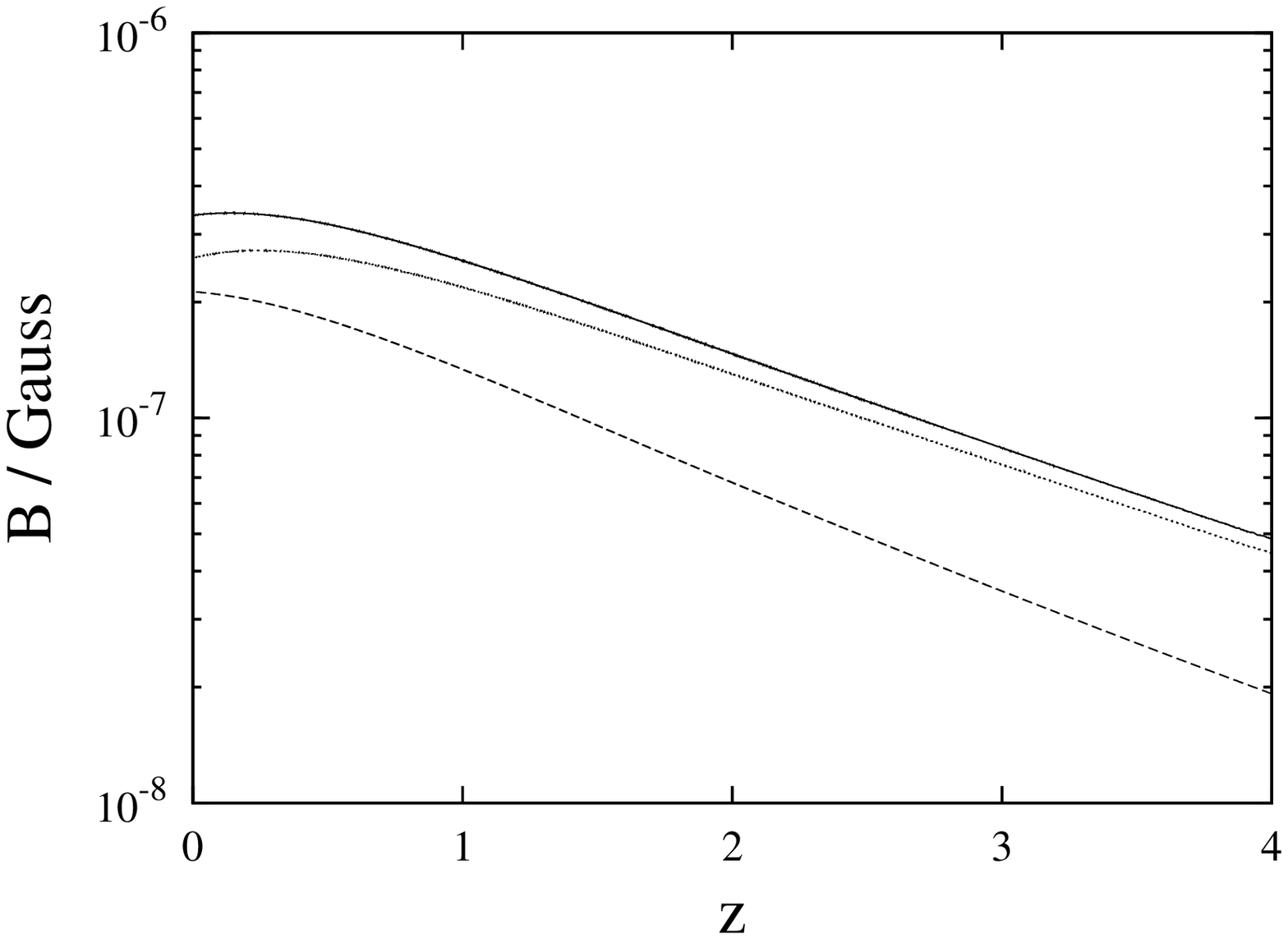}}
	\subfigure[$M_{\rm halo} = 10^{13}$~\Msun\/]{\includegraphics[height=0.45\textwidth,clip,angle=270]{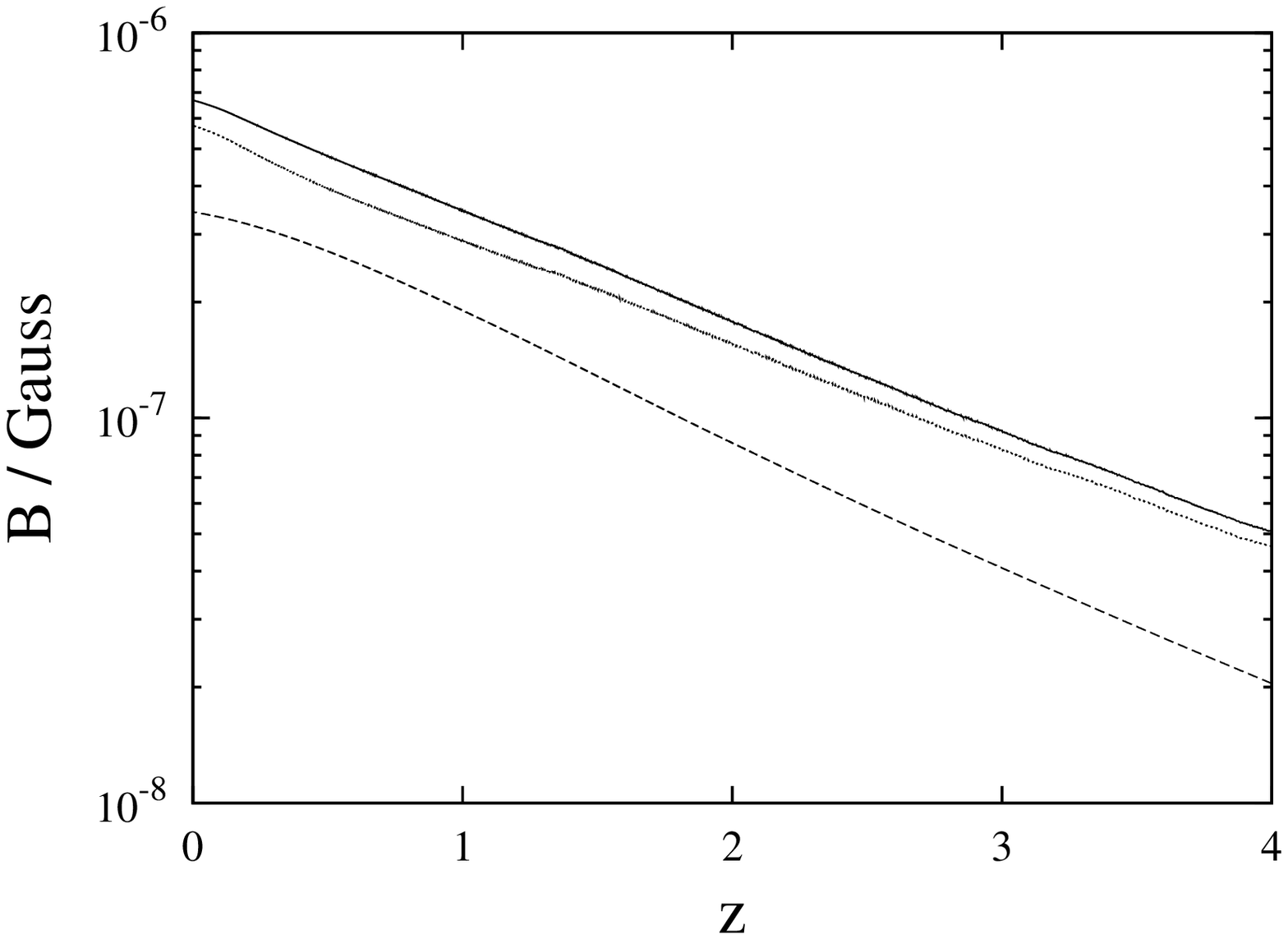}}
	\subfigure[$M_{\rm halo} = 10^{15}$~\Msun\/]{\includegraphics[height=0.45\textwidth,clip,angle=270]{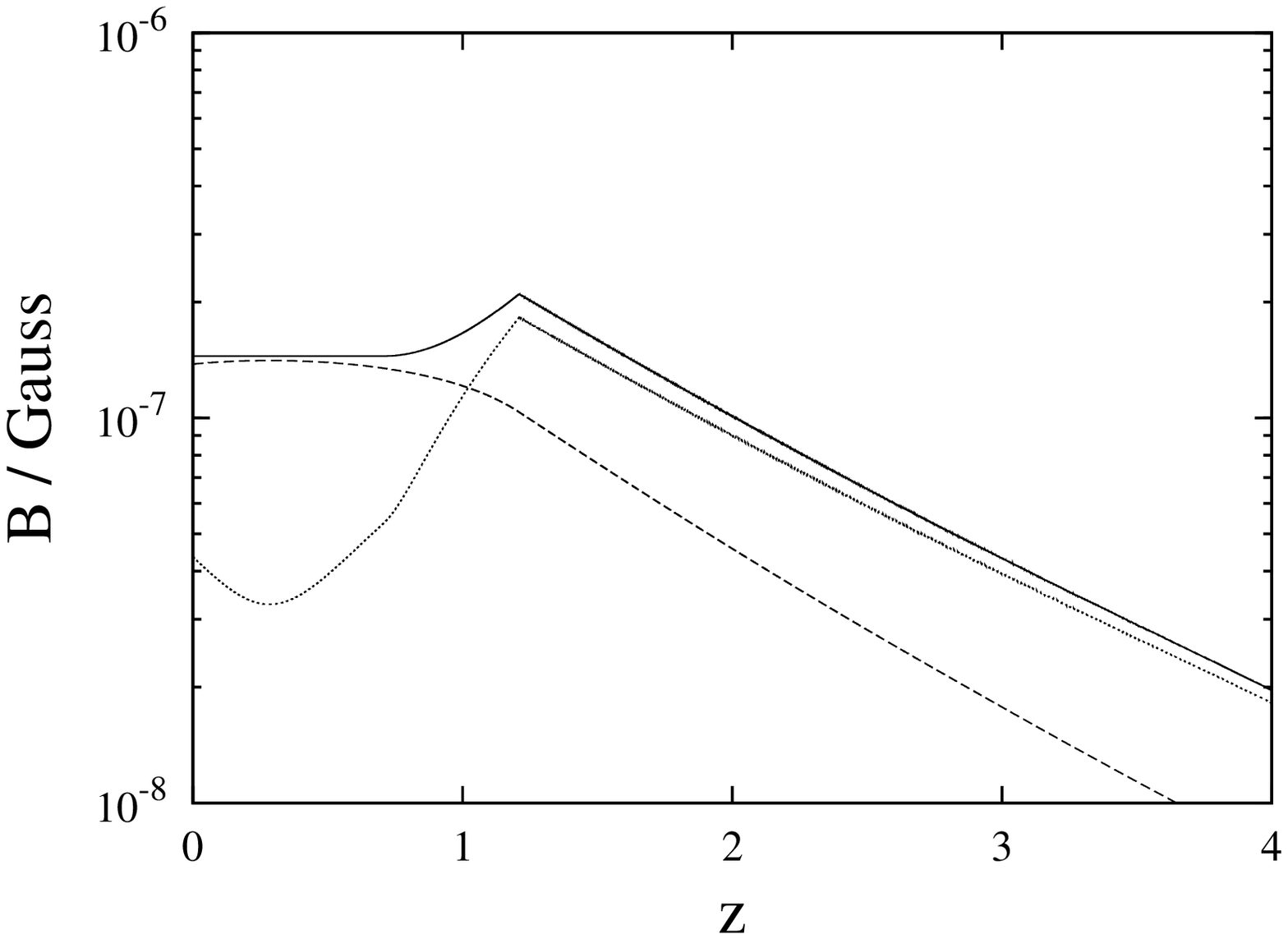}}
\caption{Contributions to the total magnetic field (solid lines) from the random (dotted) and ordered (dashed) components as a function of halo mass. All models have $\epsilon_{\rm SF}=1$, $\epsilon_{\rm SNe}=0.01$, $\epsilon_{\rm grav}=0.003$.}
\label{fig:BcomponentsVsZ} 
\end{figure*}

Figure~\ref{fig:BcomponentsVsZ} shows the evolution of the magnetic field for each of these haloes. The volume of the radio disk used to calculate magnetic field strengths was taken as $\pi R_{\rm disk}^2 T_{\rm disk}$, where the disk radius is $R_{\rm disk} = 0.1 R_{\rm vir}$ \cite{SA09} and we adopt disk thickness of $T_{\rm disk}=0.6$~kpc \cite{FA93}. In practice, disk radii derived from radio observations are subject to selection effects. We return to this point in the following section. Key features of the model are evident in Figure~\ref{fig:BcomponentsVsZ}. In low-mass haloes, interplay between the infall of cool gas and star formation result in substantial contributions to the turbulent energy budget right up to $z=0$ (Figure~\ref{fig:effPot}). As a result, the bulk of the turbulent (and hence magnetic field) energy density is in the random component. As halo mass increases, so does the strength of AGN feedback. Consequently, the most massive galaxies (Figure~\ref{fig:BcomponentsVsZ}d) are red and dead, with the bulk of the cooling and star formation taking place at $z \geq 1$. Thus, by $z=0$ a greater fraction of the turbulent energy has had time to be transferred to the ordered field than in their lower-mass counterparts.

\section{Magnetic fields in galaxy disks}
\label{sec:BfieldVsStellarMass}

\subsection{Disk sizes}
\label{sec:diskSizes}

So far we have assumed that radio emission is uniform across the disk. Contrary to the assumptions used to plot Figures~\ref{fig:BcomponentsVsZ} and \ref{fig:BvsMhalo}, in real disks the emissivity typically decays exponentially away from the disk centre \cite{Freeman70}. As a result, essentially all the source flux is observed within a few scale heights, while the observed radial extent of the disk will be greatly affected by selection effects such as the limiting beam-averaged surface brightness. In general, it is difficult to quantify these effects, and we choose to parametrise the size of the radio disk via the quantity $\epsilon_{\rm disk}$, such that $R_{\rm disk,radio}=\epsilon_{\rm disk} R_{\rm disk}$.

\begin{figure*}
        \centering
        \includegraphics[height=0.5\textwidth,angle=270]{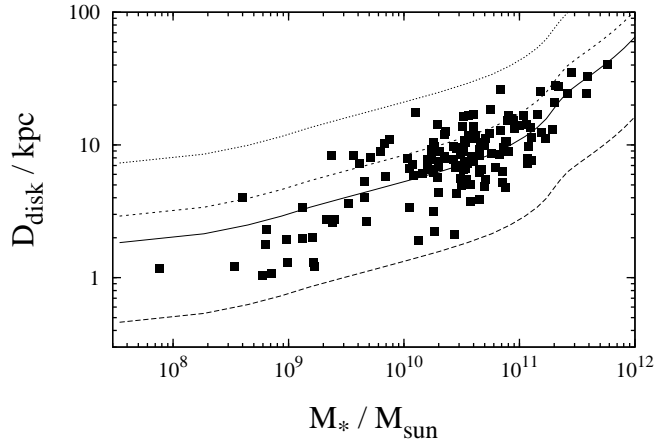}
\caption{Distribution of radio disk sizes as a function of stellar mass. Curves are for models with $\epsilon_{\rm disk}=0.06$ (long-dashed), 0.25 (solid), 0.40 (short-dashed) and 1.0 (dotted). Data points are from Fitt \& Alexander (1993).}
\label{fig:diskSizes} 
\end{figure*}

Figure~\ref{fig:diskSizes} compares predicted disk sizes with observations of late-type galaxies from the sample of Fitt \& Alexander (1993). These authors considered radio properties of an optically-complete subsample of 165 late-type galaxies brighter than magnitude $B_T = +12$, drawn from the Revised Shapley Ames Catalogue (RSA; Sandage \& Tammann 1981). These were complemented by VLA observations at 1.49 GHz \cite{Condon87}, yielding an optically-complete sample of 146 galaxies with resolved radio disks. Of the remaining 19 galaxies, one had unresolved radio stucture, eight more had no radio detections; and the remainder suffered from confusion or were not properly imaged by the observations.

Stellar masses in Figure~\ref{fig:diskSizes} and the subsequent discussion were derived from $K$-band magnitudes under the assumption that all stars are of solar type \cite{SAAR08,NikolicEA04}. Comparison of predicted and observed disk sizes shows that setting $\epsilon_{\rm disk}=0.25$ gives good agreement with observations. It is worth noting that since smaller disks are associated with higher turbulent energy densities (for fixed total turbulent energy), greater magnetic field strengths are predicted for these.

\subsection{Magnetic field strengths}
\label{sec:BfieldObservations}

\begin{figure*}
        \centering
        \includegraphics[height=0.5\textwidth,angle=270]{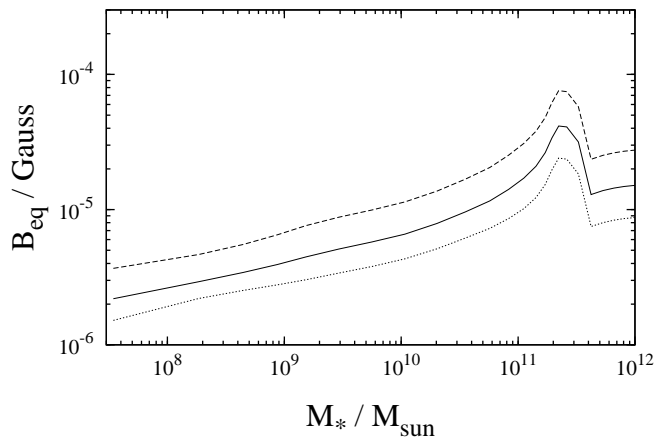}
\caption{Distribution of total magnetic field strengths $B$ as a function of stellar mass. Curves are for models with $\epsilon_{\rm grav}=0.001$ (dotted), 0.003 (solid) and 0.01 (dashed). Data points are from Fitt \& Alexander (1993) corrected for a non-negligible proton energy density (see text). Radio disk size is fixed at $R_{\rm disk,radio}=0.25 R_{\rm disk}$ as given by Figure~\ref{fig:diskSizes}. All models have $\epsilon_{\rm SF}=1$, $\epsilon_{\rm SNe}=0.01$.}
\label{fig:BvsMstars} 
\end{figure*}

Fitt \& Alexander (1993) used the observed angular sizes of the radio disk to derive equipartition magnetic field strengths. They modelled the synchrotron disk in each galaxy as having the radius given by imaging the galaxy at 1.49~GHz; and took an equivalent disk width of 0.6~kpc. This is the disk thickness we adopt for the rest of this paper. With disk volume fixed, magnetic field strength is a function of a single parameter $\epsilon_{\rm grav}$, parametrising the efficiency with which the potential energy of infalling cold gas is converted to turbulent energy. The predicted and observed magnetic field strengths are shown as a function of stellar mass for different values of $\epsilon_{\rm grav}$ in Figure~\ref{fig:BvsMstars}.

VLA sensitivity sets a limit on the strength of the weakest detectable magnetic field in the Fitt \& Alexander (1993) sample. This is given by (see Fitt \& Alexander Equation~2)
\begin{equation}
        \frac{B_{\rm min}}{\mu{\rm G}} = 384 \left[ \frac{\Sigma_{\rm min}}{\rm Jy} \left( \frac{\theta_{\rm FWHM}/2}{\rm arcsec} \right)^{-2} \right]^{2/7}
\label{eqn:FA93_Bmin}
\end{equation}
where the coefficient is different to the Fitt \& Alexander value due to their assumption of cosmic ray energy densities being negligible ($\eta_{\rm p} \sim 0$). By contrast, we assume a ratio between proton and electron energy densities of $\eta_{\rm p} = 100$ that is typical for our Galaxy (e.g. Longair 1994 and references therein), yielding an extra factor of 3.7. Beam Full Width at Half-Maximum is $\theta_{\rm FWHM} \sim 1$~arcmin, and limiting surface brightness $\Sigma_{\rm min}=0.35$~mJy. This yields $B_{\rm min} = 5.7$~$\mu$G.

\begin{figure}
        \centering
        \subfigure[$\epsilon_{\rm grav}=0.003$]{\includegraphics[height=0.45\textwidth,angle=270]{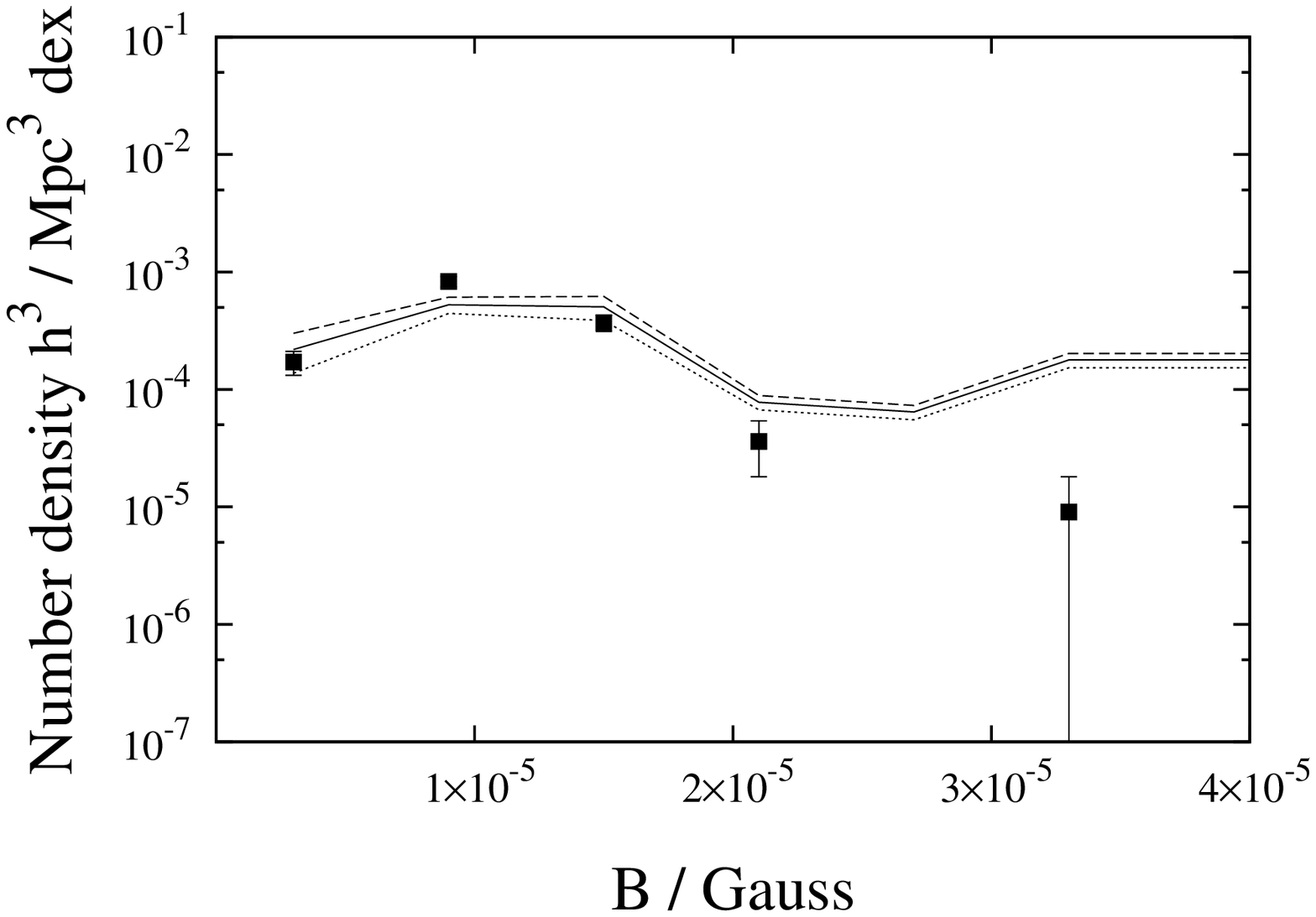}}
        \subfigure[$\epsilon_{\rm grav}=0.01$]{\includegraphics[height=0.45\textwidth,angle=270]{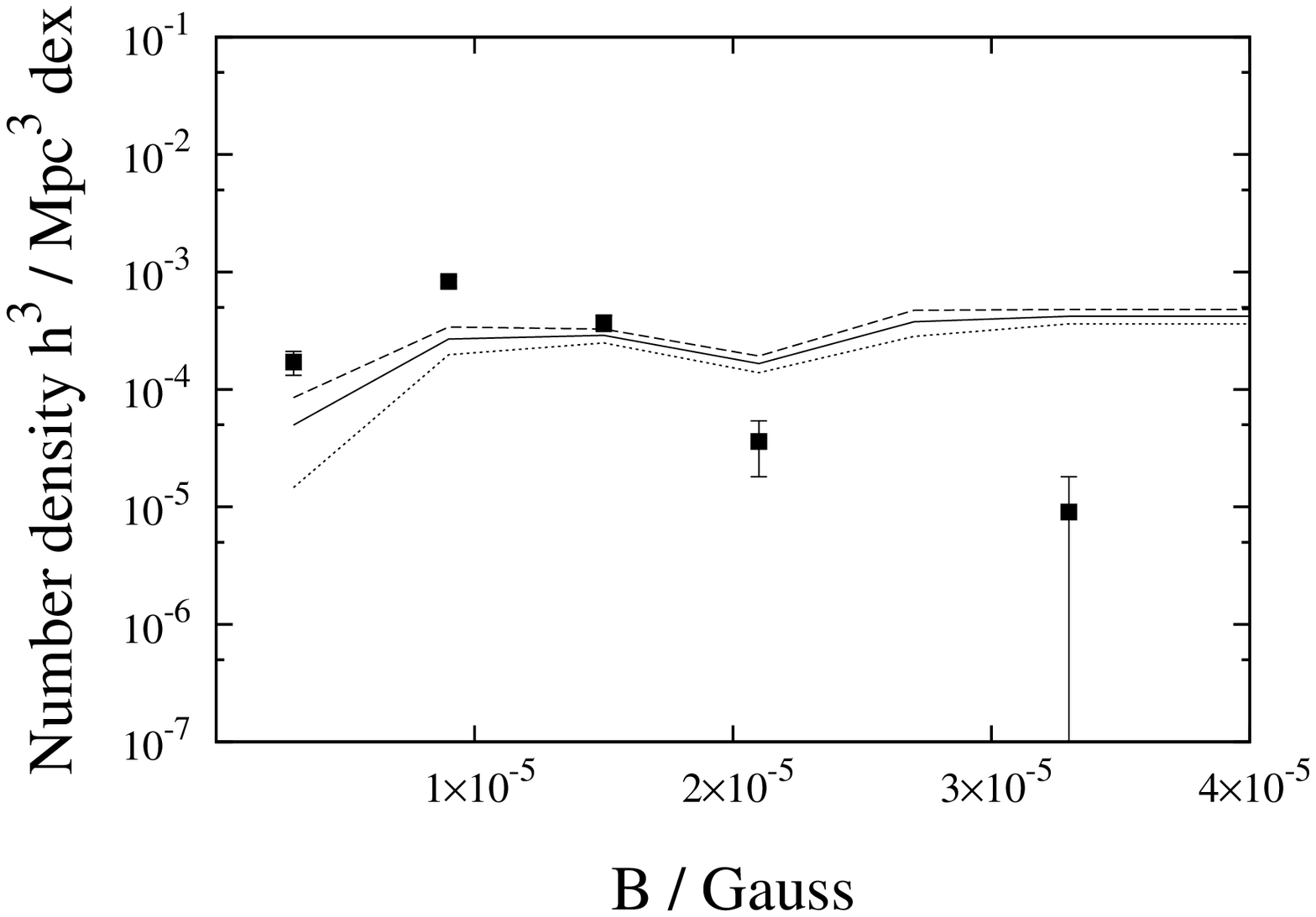}}
\caption{Distribution of total magnetic field strengths. {\it Top panel}: $\epsilon_{\rm grav}=0.003$. {\it Bottom panel}: $\epsilon_{\rm grav}=0.01$. Data points are again from Fitt \& Alexander (1993). The vertical solid line shows the minimum field strength required for the object to have detectable radio emission. Observed points below this value correspond to RSA galaxies with no radio identifications. Model curves show the mean predicted value (solid line) and upper and lower limits (dashed lines) due to uncertainties in the observed stellar mass function.}
\label{fig:Bdist} 
\end{figure}

Figure~\ref{fig:Bdist} plots the distribution of magnetic field strengths. These were obtained by binning the Fitt \& Alexander galaxies in stellar mass, and convolving this distribution with the results of Figure~\ref{fig:BvsMstars}. Our model (left panel) provides a good match to the data at low and intermediate magnetic field strengths, corresponding to low and intermediate mass galaxies. The model overpredicts the counts at the bright end, primarily due to the expected peak in $B$ at \Mstar\/$\sim 2 \times 10^{11}$~\Msun\/ (Figure~\ref{fig:BvsMstars}). A model with $\epsilon_{\rm grav}$ a factor of 3 higher than in our best fit model is included for comparison (right panel). Counts at low values of $B$ are significantly underpredicted, consistent with this model predicting higher values of $B$ at a given stellar mass than found by Fitt \& Alexander (Figure~\ref{fig:BvsMstars}).

\section{Local Radio Luminosity Function in late-type galaxies}
\label{sec:localRLF}

\subsection{Synchrotron luminosity}
\label{sec:synchLuminosity}

Synchrotron luminosity is the most straightforward observational manifestation of magnetic fields. Following Longair (1994), assuming equipartition this luminosity is given by

\begin{equation}
  L_{\rm synch}=\frac{2}{3 \mu_0} \frac{B^{7/2} V}{G(\alpha) \eta_{\rm p}}
\label{eqn:Leq}
\end{equation}
where $V$ is the volume of the emitting region and $\eta_{\rm p}$ the ratio between proton and electron energy densities. Consistent with observations of our Galaxy (Longair 1994 and references therein) we adopt $\eta_{\rm p} = 100$. The constant $G(\alpha)$ depends weakly on spectral index $\alpha$, and the minimum and maximum energy cutoffs $\nu_{\rm min}$ and $\nu_{\rm max}$. For an electron power-law distribution $N(E) \propto E^{-p}$ where $p=2.5$, we have $\alpha=(p-1)/2=0.75$. Adopting $\nu_{\rm min}=10$~MHz, we have (Longair 1994)

\begin{equation}
  G(\alpha)=4.15 \times 10^{11} \left( \frac{\nu}{\rm GHz} \right)^{0.75}
\label{eqn:Galpha}
\end{equation}

\subsection{Radio luminosities}
\label{sec:Lradio}

\begin{figure}
        \centering
        \includegraphics[height=0.45\textwidth,angle=270]{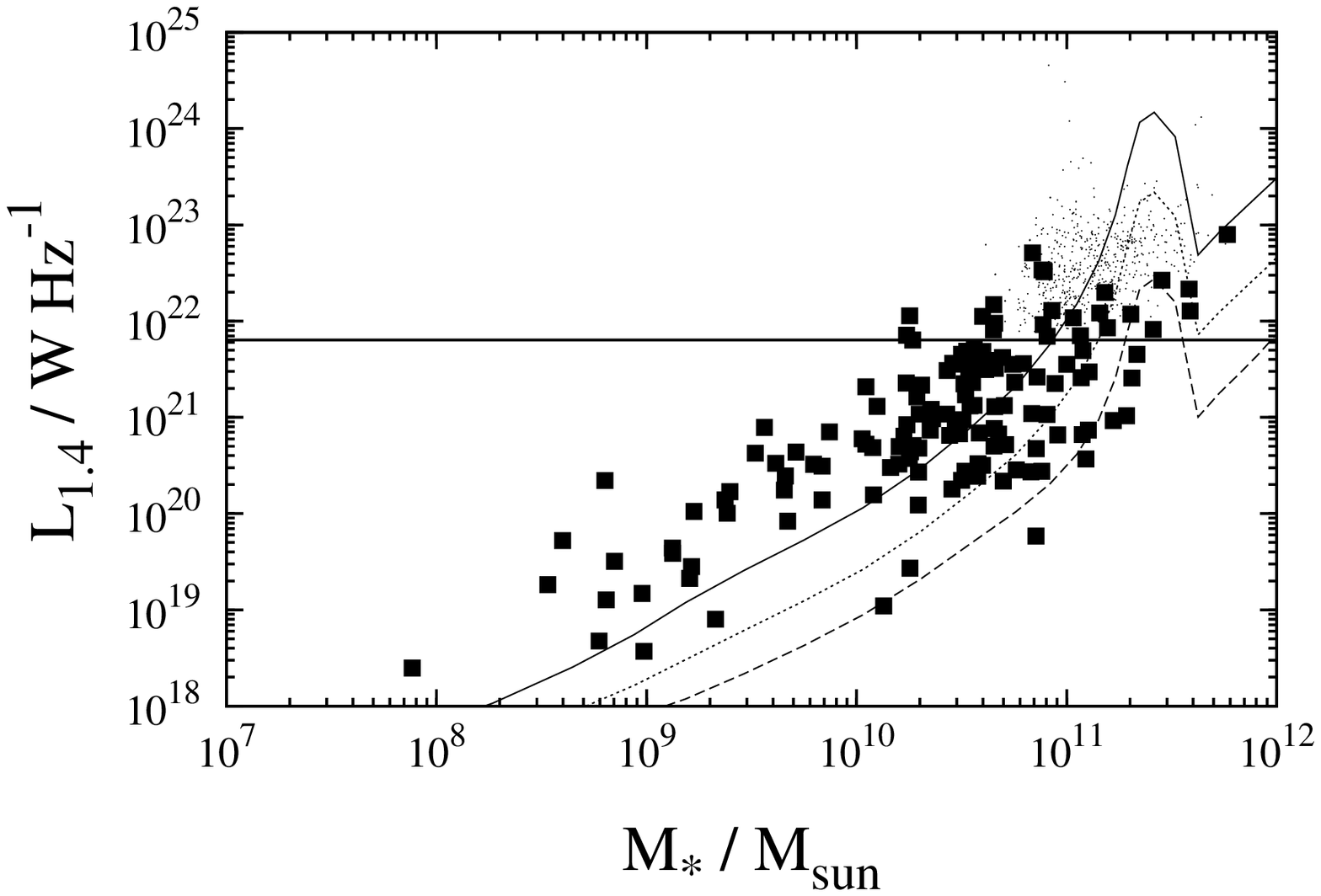}
	\includegraphics[height=0.45\textwidth,angle=270]{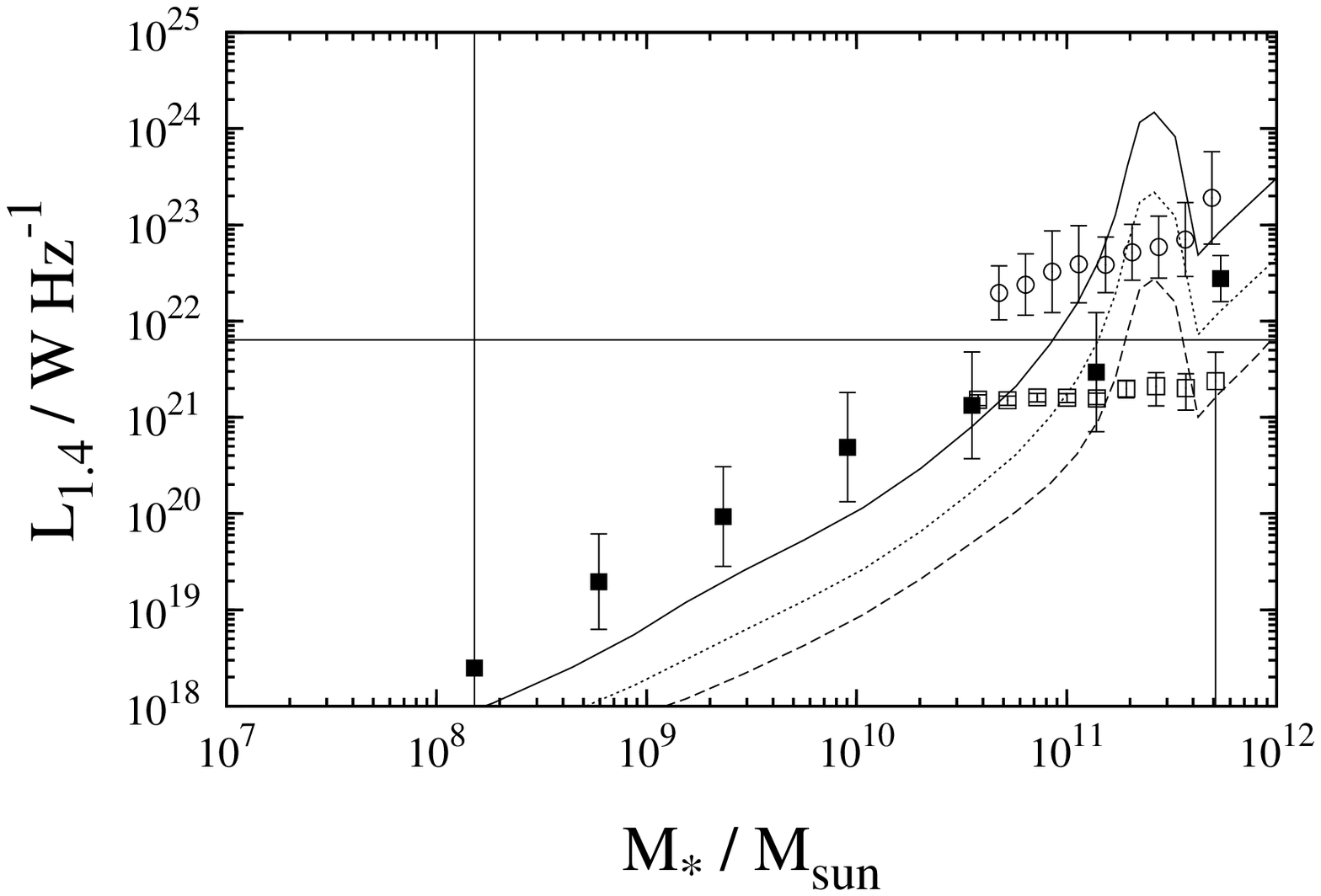}
\caption{Radio luminosity as a function of stellar mass. {\it Top panel}: All disk-type galaxies with radio detections. Data are from Fitt \& Alexander (1993; filled squares) and Shabala \etal\/ (2008; dots). The Fitt \& Alexander observations at 1.49~GHz were converted to 1.4~GHz values assuming a spectral index of 0.8. {\it Bottom panel}: Mean radio luminosity. Filled squares are again for the Fitt \& Alexander sample; open circles denote the Shabala \etal\/ (2008) sample. The Shabala \etal\/ sample is flux- and volume-limited, thus V/V$_{\rm max}$ corrections must be applied. The resultant corrected mean luminosities are shown by open squares. The minimum detectable luminosity at the front of the volume ($\sim 6 \times 10^{21}$~W\,Hz$^{-1}$ at $z=0.1$) is shown by a horizontal black line. No such corrections are required for the surface-brightness limited sample of Fitt \& Alexander (1993). Curves shows model predictions for $\epsilon_{\rm grav}=0.003$ (solid), 0.001 (dotted) and 0.0003 (dashed).}
\label{fig:LradioVsMstars} 
\end{figure}

The Fitt \& Alexander (1993) sample used in the preceding section is representative, but by no means complete. Using radio luminosity as a proxy for magnetic field strength allows us to use a complete (in both optical and radio) sample of Shabala \etal\/ (2008). Synchrotron luminosity for late-type galaxies in the sample is plotted as a function of stellar mass in Figure~\ref{fig:LradioVsMstars}. Here, late-type galaxies are defined as galaxies with concentration indices in the Petrosian $r$-band of $C>0.375$. Stellar masses were obtained from $z$-band magnitudes by assuming solar-type stars \cite{SAAR08,NikolicEA04}.

Given a disk size (i.e. the parameter $\epsilon_{\rm disk}$, set to $0.25$ in Section~\ref{sec:diskSizes}), the model predicts a single radio luminosity for each halo, and hence a given stellar mass. Figure~\ref{fig:LradioVsMstars} shows that these agree well with mean observed luminosities at low stellar masses. However, at higher masses the radio luminosity is overpredicted. The radio-loud galaxies in the flux-limited sample of Shabala \etal\/ make up the bright tail of the radio luminosity distribution. The majority of galaxies have radio luminosities below the detection limit, and thus the mean radio luminosity at a given stellar mass was calculated by assuming a Gaussian distribution with the observed scatter of $0.3$~dex. By constrast, the Fitt \& Alexander observations were performed with much greater sensitivity, and thus no such corrections are necessary.

The peak radio luminosity in Figure~\ref{fig:LradioVsMstars} corresponds to the $B$-field peak in Figure~\ref{fig:BvsMhalo}, and is related to an increase in gas cooling (and hence the random component of the magnetic field) at late times in haloes with $M_{\rm halo} \sim 10^{13}$~\Msun\/ (Figures~\ref{fig:effPot}e and f, and Figure~\ref{fig:BcomponentsVsZ}c). This late cooling is a result of the way AGN feedback is implemented in our model. In haloes more massive that $\sim 10^{12}$~\Msun\/ supernovae feedback is too weak to quench gas cooling and star formation. AGN heating can in principle do this, however only the dense central regions can be heated by weak AGNs found in haloes with mass $\leq 10^{13}$~\Msun\/. These cores have short cooling times, and thus the net effect of AGN heating is negligible. By contrast, in the most massive hosts powerful AGNs can heat rarefied gas at larger radii, where the cooling times are longer, and thus AGN heating is more effective.

As pointed out by Shabala \& Alexander (2009), an important feedback ingredient missing in our model is gas ejection by powerful radio sources. In addition to shock heating, jet-inflated radio cocoons sweep up the ambient gas through which they expand, transporting it to larger radii (e.g. Alexander 2002; Basson \& Alexander 2003). The longer cooling times will then prevent the hot gas from cooling, decreasing the turbulent energy densities, magnetic field strengths and radio luminosities in AGN hosts. A comparison of the present day cold gas mass function predicted by our model with the observed H\,I mass function \cite{ZwaanEA03} indicates that the amount of cold gas in massive hosts is overpredicted by $0.5-1$ dex. The effects of adding AGN gas uplifting to our model on the predicted magnetic fields can be estimated by decreasing the parameter $\epsilon_{\rm grav}$ by this factor. Figure~\ref{fig:LradioVsMstars} shows that this provides a better fit to observations, however the rise in radio luminosity between $\Mstar\/ = 10^{12}-3 \times 10^{12}$~\Msun\/ is still too steep. This is not surprising, since this is precisely the range where distributed supernovae feedback cannot stop large amounts of gas from cooling, while AGN heating is not yet powerful enough to affect gas at large radii with long cooling times. One would expect the gas uplifting to play a more important role in these intermediate-mass haloes.

\subsection{Radio luminosity function}
\label{sec:RLF}

The stellar mass function for local late-type galaxies was constructed from the Shabala \etal\/ sample. This stellar mass function was then converted into a halo mass function for late-type galaxies by using the best-fit model of Shabala \& Alexander (2009). The radio luminosity function can then be constructed from the halo mass function and predictions for individual haloes (Figure~\ref{fig:LradioVsMstars}).

\begin{figure}
        \centering
        \includegraphics[height=0.45\textwidth,angle=270]{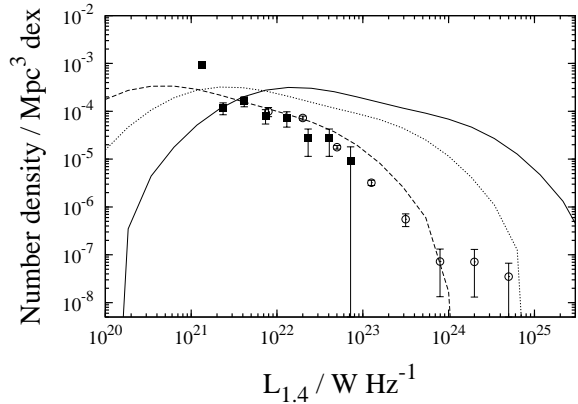}
\caption{Radio luminosity function for late-type galaxies. Model predictions are again given for $\epsilon_{\rm grav}=0.003$ (solid curve), 0.001 (dotted) and 0.0003 (dashed). Observed points are from Fitt \& Alexander (1993; filled squares) and Shabala \etal\/ (2008; open circles). V/V$_{\rm max}$ corrections are once again applied to the Shabala \etal\/ sample.}
\label{fig:RLF} 
\end{figure}

The resultant RLF is shown in Figure~\ref{fig:RLF}. As in Figure~\ref{fig:LradioVsMstars}, the number of bright sources is vastly overpredicted. Decreasing the amount of cold gas by a factor $\sim 10$ reconciles the model with observations, suggesting outward gas transport by powerful radio sources is important to the evolution of galactic magnetic fields. Observationally it should be possible to distinguish between the gas heating versus ejection scenarios through use of X-ray data. Frequent AGN heating and subsequent cooling (due to short cooling times) in dense central regions would manifest itself in copious amounts of X-ray emission. By contrast, gas transport to large radii would simply suppress cooling, and thus not give rise to any appreciable X-ray signatures.

\section{Summary}
\label{sec:conclusions}

We have developed an analytical model to follow the cosmological evolution of magnetic fields in disk galaxies. This is done by assuming equality between the magnetic field energy density, and the turbulent energy density in cold disk gas. The gas cooling and star formation histories are followed using the Shabala \& Alexander (2009) galaxy evolution model, which includes a physically-motivated prescription for AGN feedback, together with supernovae and reionization feedback. The model successfully reproduces the observed stellar mass functions and shutdown of star formation in massive galaxies out to redshifts of $1.5$.

Three mechanisms alter the turbulent energy budget in the cold gas. Star formation removes turbulence associated with gas parcels that collapse to form the stars. The turbulence is increased by gravitational infall of cool gas onto the disk; and also by high-mass stars driving shocks through the ISM as they end their lives in supernovae explosions. Two types of magnetic fields are identified: the random field, which is incoherent on scales comparable with the turbulent eddy scale; and the ordered field, generated from the random field by differential rotation of the galaxy.

Two local samples are used to test the models. The model reproduces magnetic field strengths and radio luminosities well across a wide range of low and intermediate-mass galaxies. However, the radio luminosities and magnetic field strenths are overpredicted significantly in high mass galaxies due to an overprediction in the amount of gas cooling. Inclusion of outward gas transport by powerful radio sources is required to reconcile the model with observations.

\section*{Acknowledgements}

This work has made use of the Distributed Computation Grid of the University of Cambridge (CamGRID). SS is grateful to New College, Oxford for a Research Fellowship. JMGM would like to thank STFC for a postgraduate award.

\end{document}